\documentclass{ws-mpla}
\usepackage[super,compress]{cite}
\usepackage{graphicx}
\usepackage{multirow}
\usepackage{hyperref}
\newcommand{\pt}{$p_{T}$}
\newcommand{\ttm}{$\tau\to3\mu$}

\begin{document}

\markboth{Jian Wang}{Searching for Lepton Flavor Violation at the CMS Experiment}

\catchline{}{}{}{}{}

\title{Searching for Lepton Flavor Violation with the CMS Experiment
}

\author{Jian Wang}
\address{Department of Physics, University of Florida\\
Gainesville, FL 32611-8440, USA\\
jian.wang@cern.ch}

\maketitle

\pub{Received (Day Month Year)}{Revised (Day Month Year)}

\begin{abstract}
Searches for lepton flavor violation (LFV) stand at the forefront of experimental particle physics research, offering a sensitive probe to many scenarios of physics beyond the Standard Model. 
The high proton-proton collision energy and luminosity provided by the CERN Large Hadron Collider (LHC) and the excellent CMS detector performance allow for an extensive program of LFV searches.
This article reviews a broad range of LFV searches conducted at the CMS experiment using data collected in LHC Run 2, including \ttm~decays, Higgs boson decays, and top quark production and decays. In each analysis, the online and offline event selections, signal modeling, background suppression and estimation, and statistical interpretation are elucidated. These searches involve various final state particles in a large transverse momentum range, showcasing the capability of the CMS experiment in exploring fundamental questions in particle physics.

\end{abstract}

\ccode{PACS Nos.: include PACS Nos.}

\section{Introduction}	

Lepton flavor is conserved in the Standard Model (SM) of particle physics. That is to say, electrons ($e^-$) and electron neutrinos ($\nu_e$) carry electronic lepton numbers ($L_e$) of +1, their anti-particles ($e^+$ or $\Bar{\nu_e}$) of -1, and the total $L_e$ does not change in any interaction. The same is true for the muonic and tauonic lepton numbers ($L_{\mu}$ and $L_{\tau}$). The decay $\mu^- \to e^- \Bar{\nu_e} \nu_{\mu}$ is allowed, while the lepton flavor violating (LFV) decay $\tau^- \to \mu^- \gamma$ is not.

\begin{figure}[hbtp]
\centerline{
\includegraphics[width=3.0in]{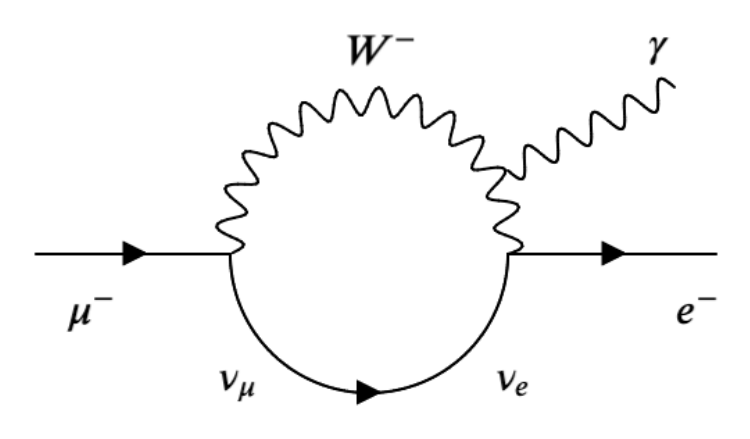}
}
\vspace*{8pt}
\caption{A Feynman diagram contributing to the $\mu \to e \gamma$ decay through neutrino oscillations.
\protect\label{feynman}}
\end{figure}

The observation of neutrino oscillations indicates that neutrinos are not massless, and LFV in the charged-lepton sector is expected to be induced by neutrino oscillations through diagrams such as that in Fig.~\ref{feynman}. The branching fraction of such a decay is expected to be $\lesssim O(10^{-50})$, which is completely negligible for all practical purposes.  Searching for LFV processes is therefore free of SM background, and ideal to probe beyond the SM (BSM) physics. Indeed, several BSM models hypothesize LFV decays at rates that are measurable at present or near-future experiments. A detailed description of these models can be found in Ref.~\refcite{Calibbi:2017uvl}).

Experimental explorations of LFV started in 1940s when B. Pontecorvo et al. searched for $\mu \to e\gamma$ decays using cosmic muons~\cite{Hincks:1948vr}. Experiments using pion beams, and later muon beams, were performed in the decades that followed. It is worth noting that the early null results on muon LFV decays provided indirect evidence that muon neutrinos and electron neutrinos were different particles. The current best upper limits on the branching fractions of these decays are $O(10^{-13}-10^{-12})$~\cite{SINDRUM:1987nra,MEG:2016leq,MEGII:2023ltw}.

Tau-lepton ($\tau$) LFV decays ($\tau \to \mu \gamma$, $\tau \to \mu ee$, etc) have been searched for since 1980s at electron-positron colliders ( $e^+ e^- \to \tau^+\tau^-$). The current best upper limits on the branching fraction of these processes are typically $O(10^{-8})$~\cite{BaBar:2009hkt,Belle:2021ysv,Hayasaka:2010np}. 
The relations between muon LFV decays and $\tau$ LFV decays are model dependent such that the extremely small limits on the former do not demotivate the studies of the latter.

Since the start of its operations in 2010, the CERN Large Hadron Collider (LHC) has provided another prolific source of $\tau$ leptons via proton-proton (pp) collisions. At the same time, the LHC offers a unique opportunity to access LFV processes involving heavy particles, such as the Z boson, the Higgs boson, or the top quark. The CMS experiment at the LHC has recorded $5~fb^{-1}$ and $20~fb^{-1}$ of pp collision data, at center-of-mass energies of 7~TeV and 8~TeV, respectively, during the 2010-2012 data taking period (known as "Run 1"), and $138~fb^{-1}$ of pp collision data at a center-of-mass energy of 13~TeV during the 2016-2018 data taking period (known as "Run 2"). The studies reviewed in this article are performed using the Run 2 data.

The review is organized as follows. Section~\ref{sec:cms} gives a brief overview of the experimental apparatus. Sections~\ref{sec:tau},~\ref{sec:higgs}, and~\ref{sec:top} describe the searches for LFV in \ttm~decays, in Higgs boson decays, and in top quark productions and decays, respectively. A summary is provided in Section~\ref{sec:summary}. Searches for heavy BSM particles that involve LFV decays, e.g. $Z' \to e\mu$, are not in the scope of this review.

\section{The CMS detector }
\label{sec:cms}

The central feature of the CMS apparatus is a superconducting solenoid of 6 m internal diameter, providing a magnetic field of 3.8 T. Within the solenoid volume are a silicon pixel and strip tracker, a lead tungstate crystal electromagnetic calorimeter (ECAL), and a brass and scintillator hadron calorimeter (HCAL), each composed of a barrel and two endcap sections. Forward calorimeters extend the pseudorapidity coverage provided by the barrel and endcap detectors. 
Outside the solenoid are gas-ionization muon detectors embedded in the steel flux-return yoke.
A more detailed description of the CMS detector, together with a definition of the coordinate system used and the relevant kinematic variables, can be found in Ref.~\refcite{CMS:2008xjf}.

Events of interest are selected using a two-tiered trigger system. The first level (L1), composed of custom hardware processors, uses information from the calorimeters and muon detectors to select events at a rate of around 100~kHz within a fixed latency of $4\mu s$~\cite{CMS:2020cmk}. The second level, known as the high-level trigger (HLT), consists of a farm of processors running a version of the full event reconstruction software optimized for fast processing, and reduces the event rate to around 1~kHz before data storage~\cite{CMS:2016ngn}.

A global ``particle-flow'' (PF) algorithm~\cite{CMS:2017yfk} aims to reconstruct all individual particles in an event, combining information provided by the silicon tracker, ECAL, HCAL, and the muon detectors. 
Muons are measured in the pseudorapidity range $|\eta| < 2.4$, with detection planes made using three technologies: drift tubes, cathode strip chambers, and resistive plate chambers~\cite{CMS:2018rym}. Matching muons to tracks measured in the silicon tracker results in a relative transverse momentum (\pt) resolution, for muons with \pt up to 100~GeV, of 1\% in the barrel and 3\% in the endcaps. 
The electron momentum is estimated by combining the energy measurement in the ECAL with the momentum measurement in the tracker~\cite{CMS:2020uim}. The momentum resolution for electrons with \pt $\approx$ 45~GeV from $Z \to ee$ decays ranges from 1.6 to 5\%.
Hadronic $\tau$ decays ($\tau_h$) are reconstructed from jets, using the hadrons-plus-strips algorithm~\cite{CMS:2018jrd}, which combines 1 or 3 tracks with energy deposits in the calorimeters, to identify the $\tau$ decay modes. Neutral pions are reconstructed as strips with dynamic size in $\eta$-$\phi$ ($\phi$ being the azimuthal angle) from reconstructed electrons and photons, where the strip size varies as a function of the \pt of the electron or photon candidate.
Hadronic jets are clustered from the reconstructed particles using the infrared and collinear safe anti-$k_T$ algorithm~\cite{Cacciari:2008gp, Cacciari:2011ma} with a distance parameter of 0.4. Jet momentum is determined as the vectorial sum of all particle momenta in the jet, and is found from simulation to be, on average, within 5 to 10\% of the true momentum over the whole \pt spectrum and detector acceptance. 
The missing transverse momentum vector $\vec{p}^{~miss}_T$ is computed as the negative vectorial sum of the \pt of all the reconstructed particles in an event, and its magnitude is denoted as $p^{miss}_T$~\cite{CMS:2019ctu}. 

All Monte Carlo (MC) generated signal or background events are processed with a full simulation of the CMS detector response based on \textsc{Geant4}~\cite{GEANT}. The presence of pileup, additional pp collisions in the same or adjacent bunch crossings, is accounted for by overlaying each simulated event with a number of \textsc{Pythia}~\cite{PYTHIA} simulated minimum bias events.

\section{Search for \ttm~decays}
\label{sec:tau}

Muons have relatively clean signatures in the detector, making \ttm~the most, perhaps the only, feasible $\tau$ LFV decay mode to be explored at the LHC. 
The previously published upper limits on the branching fraction $\mathcal{B}(\tau\to3\mu)$ by LHC experiments are $4.6 \times 10^{-8}$ by the LHCb experiment~\cite{Aaij:2014azz}, and $3.8 \times 10^{-7}$ by the ATLAS experiment~\cite{Aad:2016wce}, using their Run 1 data.
The best upper limit to date is $2.1 \times 10^{-8}$, set by the Belle experiment~\cite{Hayasaka:2010np}, followed by an upper limit of $3.3 \times 10^{-8}$ by the BaBar experiment~\cite{Lees:2010ez}. Both Belle and BaBar are electron-position collision experiments, known as "B factories". All these limits are at 90\% Confidence Level (CL). 

The production rate of $\tau$ leptons at the LHC is high - $10^{11}$ per $fb^{-1}$, according to \textsc{Pythia} predictions. 
The dominant source of $\tau$  is heavy flavor (charm or bottom) hadron decays, referred to as HF in what follows. The leading contribution is from $D_s^+$ decays (about 69\% of the total), followed by $B^+$ and $B^0$ decays (12\% each), and $B_s^0$ and $D^+$ decays (3\% each). 
Charge-conjugated processes are implied throughout this article.
A completely different source of $\tau$ is W boson decays, the production rate of which is about 0.01\% of that from HF decays.

\begin{figure}[hbtp]
\centerline{
\includegraphics[width=2.2in]{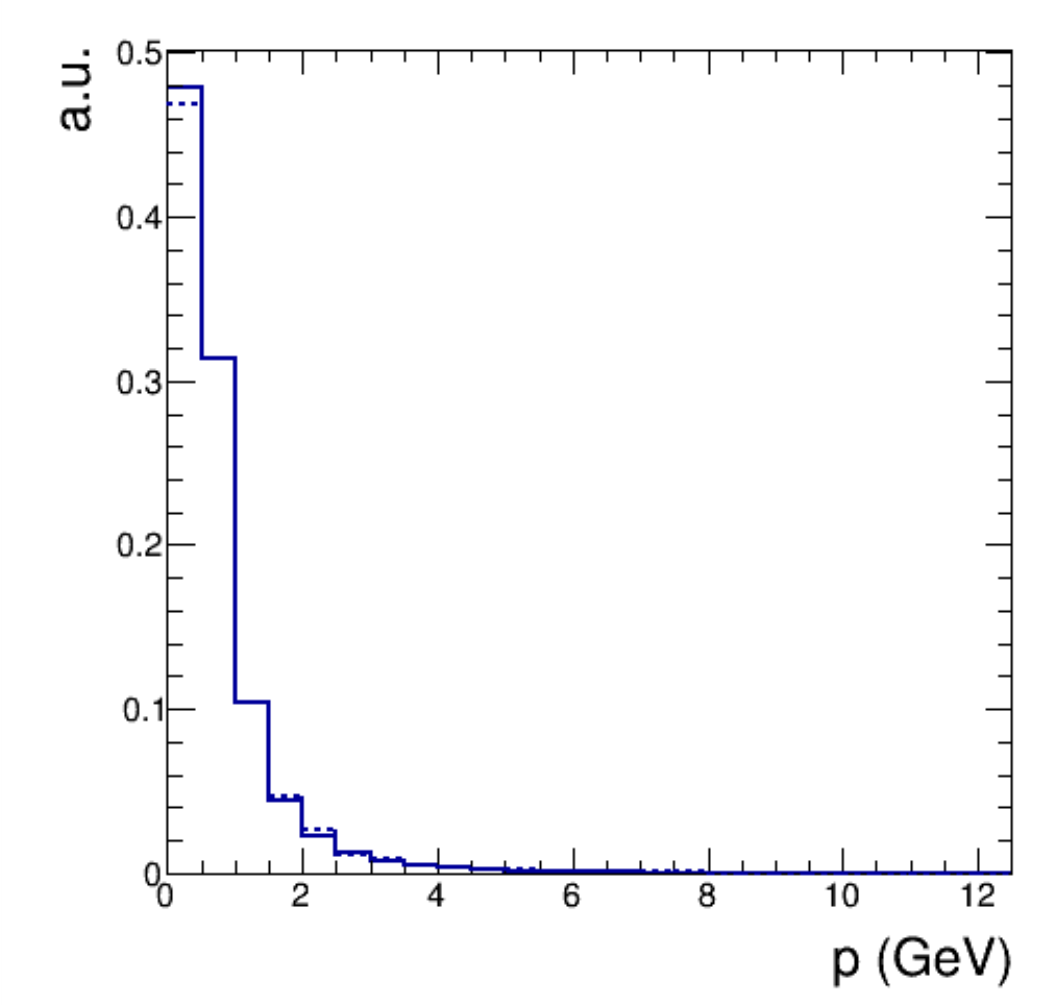}
\includegraphics[width=2.2in]{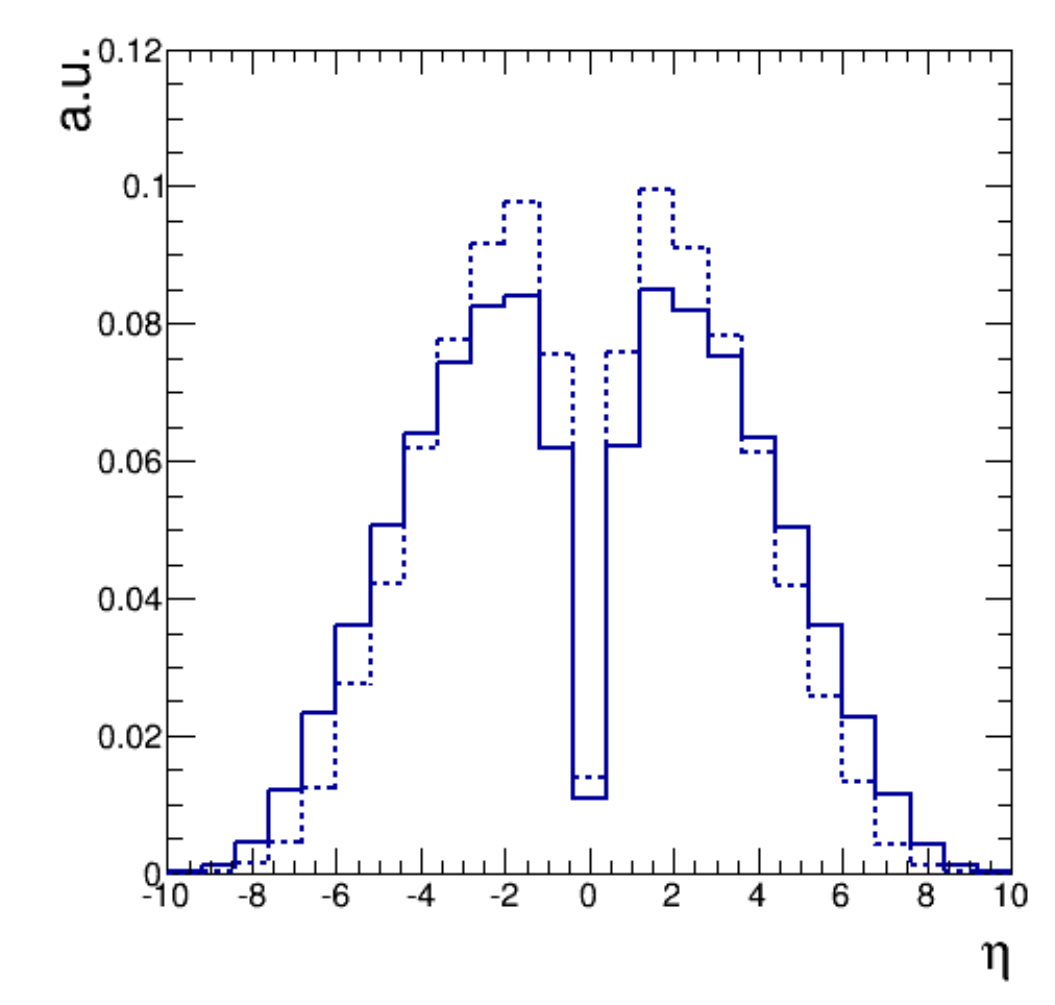}
}
\vspace*{8pt}
\caption{
Momentum ($p$) of the softest muon of the three in \ttm~decays (left).
Pseudorapidity ($\eta$) of the most forward muon of the three in \ttm~decays (right).
The solid lines indicate $\tau$ leptons from charm hadron decays, while the dashed lines indicate those from bottom hadron decays. The distributions come from \textsc{Pythia} predictions.
\protect\label{acceptance}}
\end{figure}

A muon is required to have momentum $p>3$~GeV to reach the muon detectors, as a consequence of the energy loss to penetrate the calorimeters as well as the trajectory bending in the magnetic field. The muon detection acceptance could accordingly be defined as $p>3$~GeV and $|\eta| < 2.4$. As shown in Fig.~\ref{acceptance}, the muons from HF-produced \ttm~events tend to have small $p$ and ``forward" $\eta$. In fact, only 1\% of the events would be in the acceptance. Furthermore, the L1 trigger relies on muon detectors to assign  \pt to muons. To do so, muons have to cross 3 or 4  muon detector planes (embedded in steel yoke) to ensure the quality, which implies additional momentum requirements, and consequently a further factor of 10 reduction of \ttm~events. As a result, the difference of event yields in the  HF channel and the W boson channel is about one order of magnitude instead of four.

The CMS experiment explores both HF and W boson channels to probe \ttm~decays. 
A $\tau$ lepton has a mean lifetime $t=2.9 \times 10^{-13}s$, and a mass $m=1.777$~GeV. As for a  $\tau$ lepton of $p=20$~GeV (typical $\tau$ momentum in this analysis), its decay length is $L=ctp/m\simeq1$~mm ($c$ being the speed of light). This length is big enough to be measured by the vertex detector made of silicon pixels. Therefore a \ttm~signal event, regardless of the HF source or the W boson source, features a displaced secondary vertex formed by 3 muons ("secondary" here means it is not the primary pp collision vertex). The 3 muons tend to be collimated, even more so in the W boson channel where the $\tau$ momentum is higher. 
Besides, a $\tau$ from the W boson source is isolated from hadronic activities, and accompanied by large $p^{miss}_T$ due to the neutrino from the W boson decay. It is remarkable that the CMS experiment takes advantage of the two complementary sources of $\tau$  thanks to its excellent detector performance.

\subsection{$\tau$ from heavy flavor decays}

The trigger of the HF analysis requires two or three low \pt muons, in a similar way as in other CMS heavy-flavor physics (a.k.a "B physics") analyses. The L1 trigger requires two muons with $|\eta|< 1.5$ and no explicit \pt requirement, or two muons with \pt$> 4$~GeV in the full $\eta$ coverage ($|\eta|< 2.4$). These are complemented by another L1 trigger requiring 3 muons with $p_T > 5 , 3, 0$~GeV, respectively, in the full $\eta$ coverage. A large majority of the events is collected by the two-muon trigger.

The HLT requires two muons and one track. The muons must have $p_T > 3$~GeV, and the track $p_T > 1.2$~GeV. The three objects are fitted to a common vertex, which must be displaced from the pp collision vertex by at least twice the measurement uncertainty. The invariant mass of the three objects, assuming they have muon masses, is required to be in the range of 1.6 -- 2.0 GeV. 

Three muons are required to be offline identified (by the PF algorithm). Otherwise the offline pre-selections are as loose as possible, only to consolidate the trigger selections. The three muon tracks are then refitted using the common vertex constraint~\cite{kinFit}, which improves the trimuon mass resolution by about 5\%.

\begin{figure}[hbtp]
\centerline{
\includegraphics[width=3.0in]{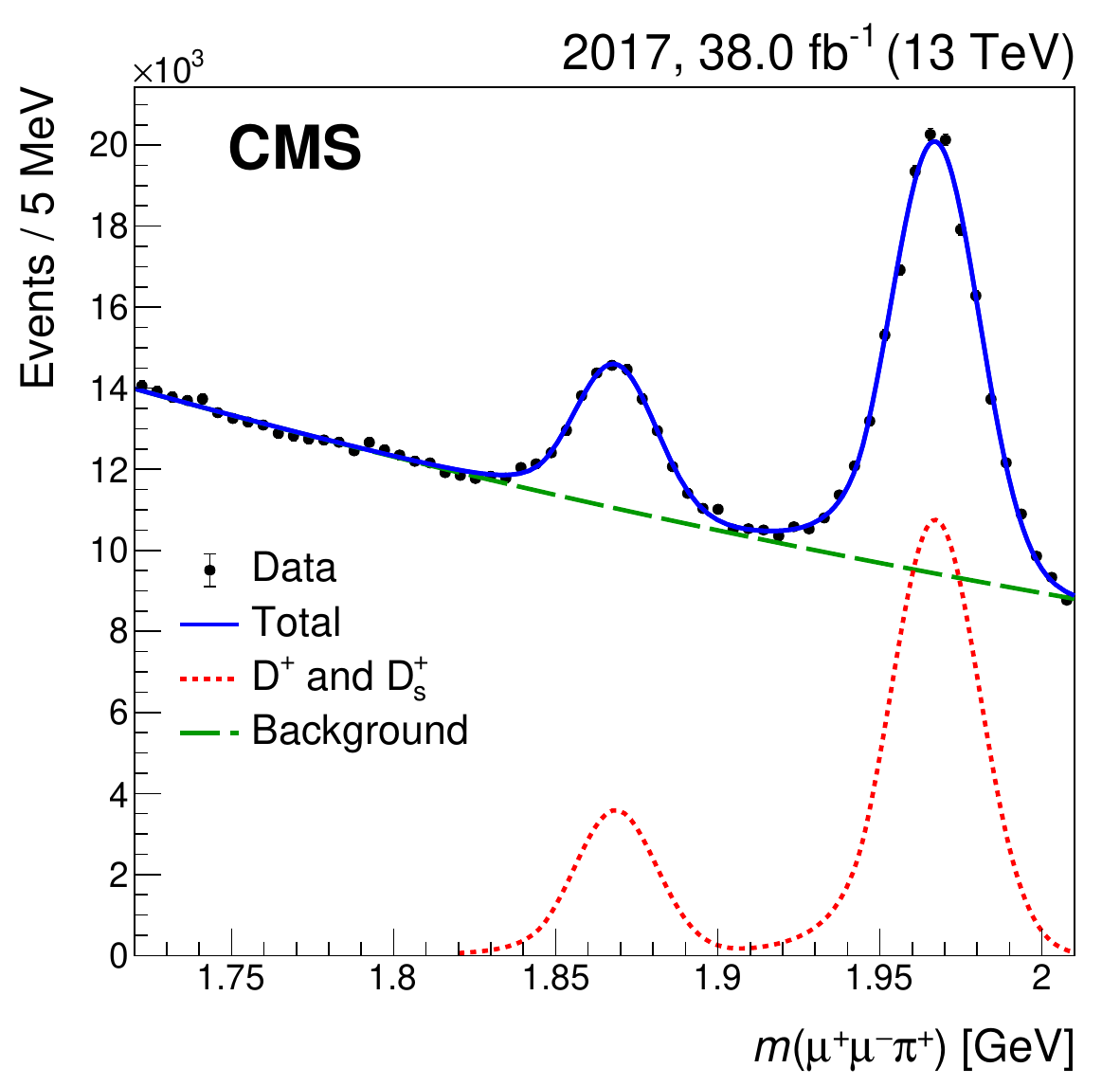}
}
\vspace*{8pt}
\caption{
The $\mu\mu\pi^+$ invariant mass distribution with the fits to the sum of the $D^+$ (1.870~GeV) and $D_s^+$ (1.968~GeV)~\cite{pdg2022} resonances and the background in 2017 data.
\protect\label{tau_normalization}}
\end{figure}

The HF analysis is complicated by the uncertainties in the bottom (b) and charm (c) hadron production rates, and also the signal acceptance and efficiencies. Selection efficiencies are usually measured per particle using a control sample as a function of relevant observables, most commonly $p_T$ and $\eta$. The measured efficiencies are then applied to the final state particles in the signal. It does not work reliably in  very low $p_T$ analyses for the reason that the $p_T$ is not in the region of the efficiency ``plateau'', which invalidates the usage of efficiencies measured from another sample. To minimize the dependence of simulation prediction, a normalization channel $D_s^+ \to \phi(\mu\mu) \pi^+$ is used.
Apart from the obvious reason that $D_s^+$ is the major source of $\tau$, $D_s^+ \to \phi(\mu\mu) \pi^+$ has very similar topology and $p_T$ distributions as the signal channel. Moreover, the $D_s^+ \to \phi(\mu\mu) \pi^+$ decay is fully reconstructable, manifesting itself as a peak in data, such that it is convenient to extract from data not only yields (in other words efficiencies), but also kinematic distributions and mass resolutions. Note that, both prompt $D_s^+$ production and $B \to D_s^+$ decays are considered in the signal channel and in the normalization channel. The aforementioned “two muons and one track” HLT  has the advantage of collecting the events of \ttm~signal and $D_s^+ \to \phi(\mu\mu) \pi^+$ normalization channel at the same time. The offline event selections on $D_s^+ \to \phi(\mu\mu) \pi^+$ are as close as possible to those for the signal events, except it has two identified muons instead of three.
The $\mu\mu\pi^+$ invariant mass distribution used to extract the $D_s^+ \to \phi(\mu\mu) \pi^+$ event yield in data is shown in Fig.~\ref{tau_normalization} (taken from Ref.~\refcite{CMS:2023iqy}).

The smaller contributions of $\tau$ from $B^+$ or $B^0$ decays are based on simulation prediction, but verified by comparing the decay length of $B \to D_s^+$ in data and simulation. The very small contributions from $B_s^0$ and $D^+$ to $\tau$ decays are taken from simulation, and 100\% uncertainties are assigned.

The dominant background is combination of two real muons and one charged hadron. A pion or a kaon has a chance to decay to a muon and a neutrino before its interaction with the calorimeters (known as “decay in flight”) - lower the momentum, higher the probability for it to occur. In some cases, especially for a kaon decay-in-flight, a "kink" could be identified in the track measured by the silicon tracker, which means the tracks have two parts that do not look compatible with each other in the direction and magnitude of the measured momentum. When no prominent kink is seen, the kaon or pion simply looks the same as a real muon would in the detector, with a high quality track in the silicon tracker, and a muon penetrating the calorimeters and registering footprints in the muon detectors.

The most common physics process mimicking the \ttm~signal is  b to c cascade decays - a b hadron decays to a c hadron, a muon, and other particles; the c hadron subsequently decays to a muon, at least one hadron, and other particles. The decay length of the c hadron has a certain probability to be so small that the two muons and one hadron seem to have emerged from the same vertex, which is displaced from the pp collision vertex as a signature of b decay. Once the hadron is misidentified as a muon, the whole event is similar in appearance to a \ttm~decay. 

There are also backgrounds with three genuine muons. To suppress them, events are vetoed if opposite-charge dimuon has an invariant mass consistent with $\phi(1020)$ or $\omega(783)$. Other specific processes, though rarer, do not have such reconstructable resonances, for example $D_s^+ \to \eta(\mu^+\mu^-\gamma) \mu^+ \nu$, and thus cannot be reduced easily.

\begin{figure}[hbtp]
\centerline{
\includegraphics[width=3.0in]{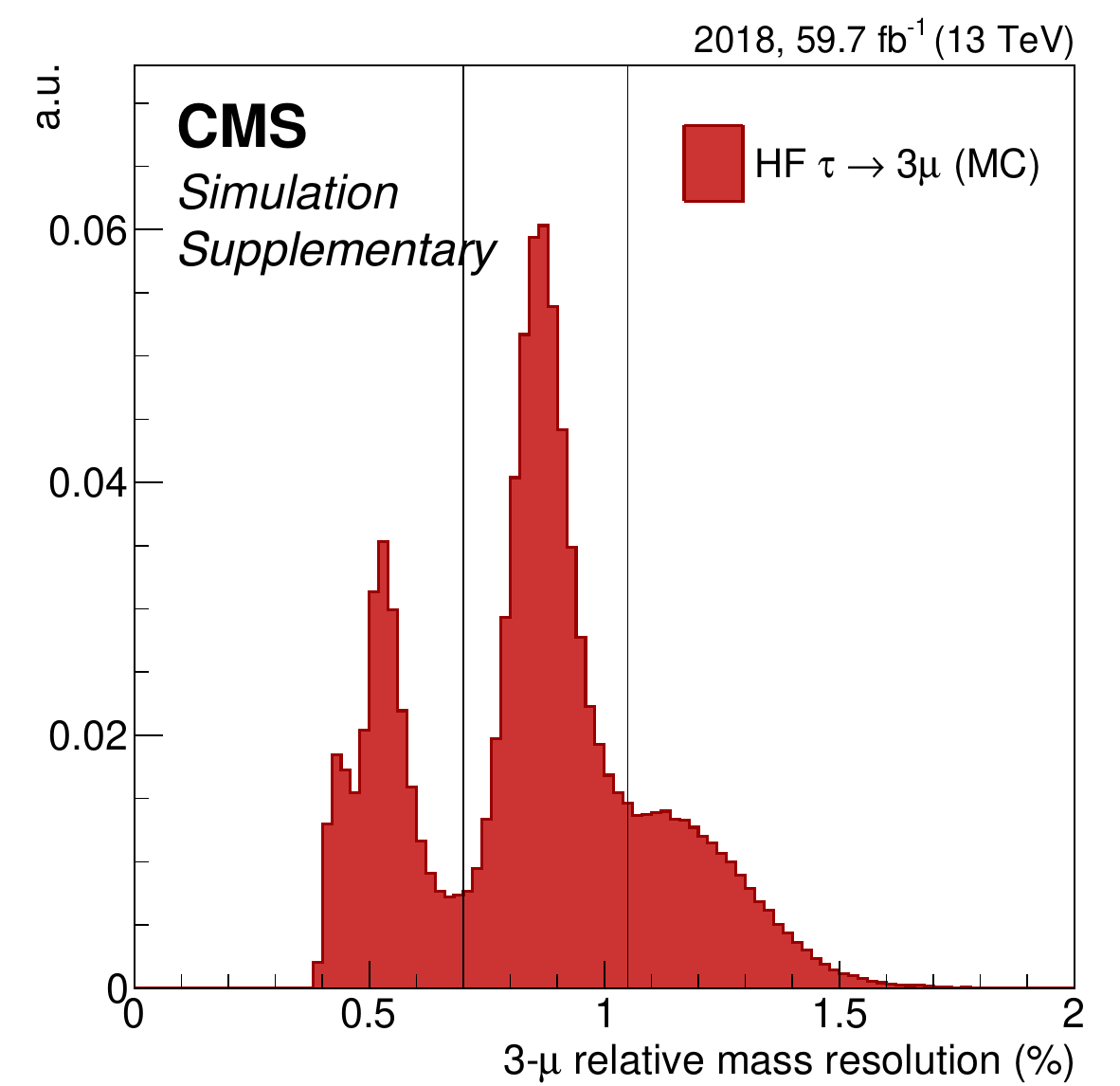}
}
\vspace*{8pt}
\caption{Relative mass resolution of trimuon candidates selected in signal MC events. The event categorisation used in the analysis is indicated by vertical lines: $< 0.7\%$ (category A), $0.7 - 1.05\%$ (category B), and $> 1.05\%$ (category C).
\protect\label{mass_resolution}}
\end{figure}

It is important to note that, there is no known peaking structure in the trimuon mass distribution 1.6--2.0 GeV, based on which the \ttm~search strategy is to look for a peak consistent with $m(\tau$) on top of a smooth background distribution. 
The trimuon mass resolution, $\sigma$, plays an important role here, simply because, better the resolution, narrower the mass peak, consequently smaller the background contamination under the signal peak. 
The track momentum resolution has a strong dependence on $\eta$, which propagates to the mass resolution. Indeed, as shown in Fig.~\ref{mass_resolution}, the relative mass resolution, $\sigma/m$, has a significant variation. The three parts separated by the vertical lines in the figure roughly correspond to the silicon tracker barrel, overlap, and endcap regions, respectively.
Events with a better mass resolution should be separated from those with a worse resolution to benefit the overall search sensitivity. Both simulated signal events and data events are therefore grouped into 3 categories, with $\sigma/m < 0.7\%, 0.7\% < \sigma/m < 1.05\%$, and $\sigma/m > 1.05\%$, denoted as categories A, B, and C, respectively. 
In each of the category, the mass range of [$m(\tau$)-2$\sigma$, $m(\tau$)+2$\sigma$] is defined as the signal region, which is not looked at until the analysis selection criteria are fully established, to avoid possible biases. On the other hand, those in 1.6--2.0~GeV but not in the signal region are called sidebands, and data events in the sidebands are used as a proxy for the background in the analysis.

\begin{figure}[hbtp]
\centerline{
\includegraphics[width=3.0in]{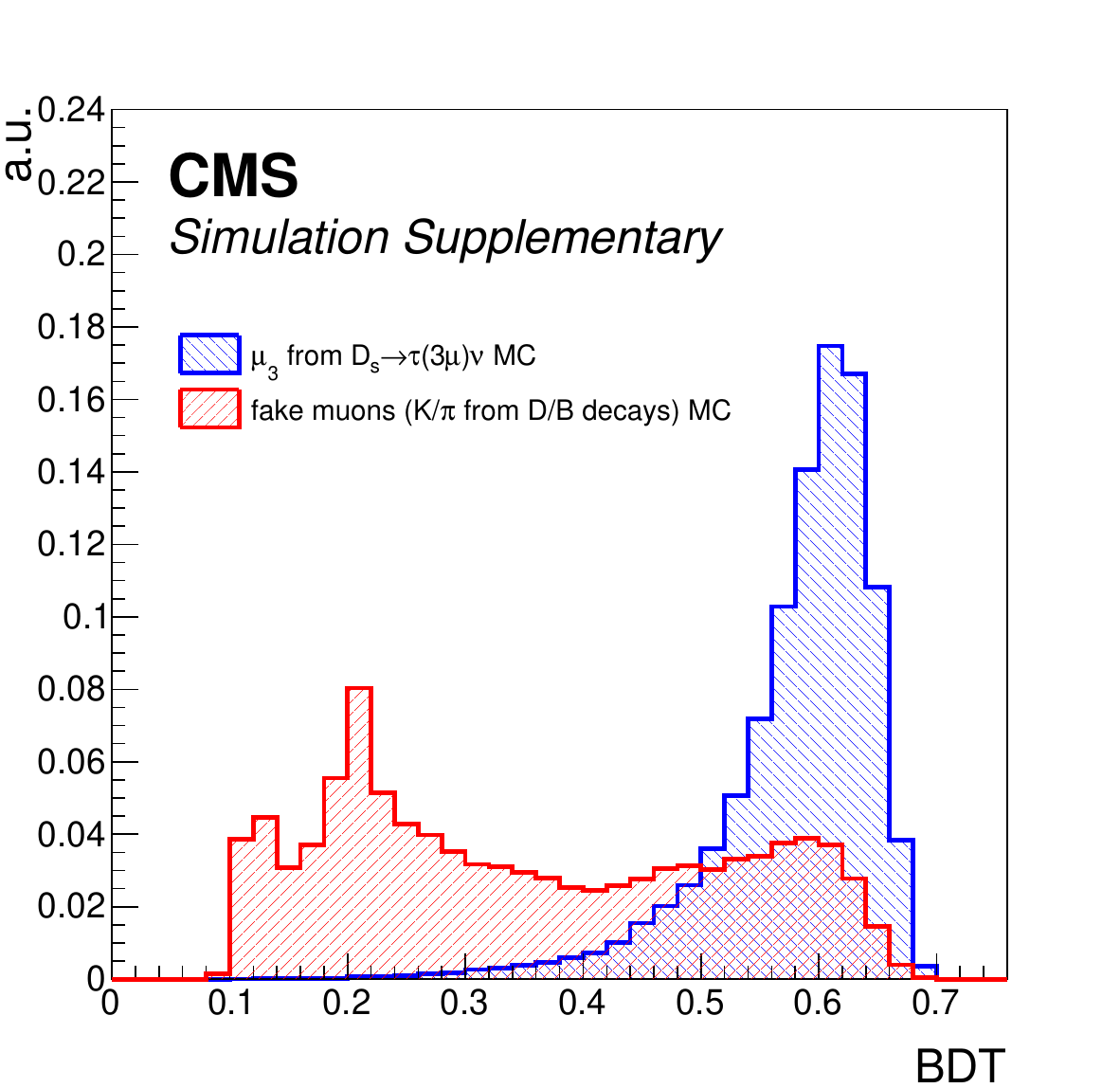}
}
\vspace*{8pt}
\caption{Distribution of muon reconstruction quality BDT score for the lowest-$p_T$
muon in signal MC (blue) and for simulated kaons or pions from D and B meson decays misidentified as muons (red). Distributions are normalised to unity. 
\protect\label{muon_BDT}}
\end{figure}

A Boosted Decision Tree (BDT) is trained~\cite{Chen:2016btl} based on muon reconstruction quality in order to distinguish genuine muons from misidentified muons. The training makes use of simulated samples, taking misidentified muons from simulated B or D meson decaying to hadrons. The input observables are silicon tracker measurements (including the aforementioned “kink”), muon detector measurements, as well as the compatibility between silicon tracker and muon detector measurements. 
Fig.~\ref{muon_BDT} shows the muon quality BDT outputs for real and misidentified muons.
The BDT is then applied to all three muons in the simulated \ttm~events and also the data sideband events. 

Another BDT (referred to as the “analysis BDT”) is trained in each mass-resolution category to separate the simulated \ttm~signal events and the data sideband events. The outputs of the muon quality BDT are taken as inputs to this analysis BDT. Other input observables are mostly characteristics of the displaced trimuon vertex, the most discriminating ones being:
\begin{itemlist}
 \item The trimuon vertex fit $\chi^2$ per degree of freedom as an indication of the vertex quality - on average larger in background.
 \item The negative of the trimuon momentum vector is expected to point back to the pp collision vertex. This is particularly true in events of $\tau$ originating from a promptly-produced $D_s^+$ meson, given the very small mass difference of $D_s^+$ and $\tau$. The pointing angle, $\alpha$, defined as the angle between the trimuon momentum vector and the line connecting the pp collision vertex and the trimuon vertex tends to be smaller in the signal.
 \item The minimum distance of extra tracks to the trimuon vertex, as an indication of additional particles produced in b to c cascade decays, tends to be smaller in background.
\end{itemlist}

\begin{figure}[hbtp]
    \centering
    \includegraphics[width=2.2in]{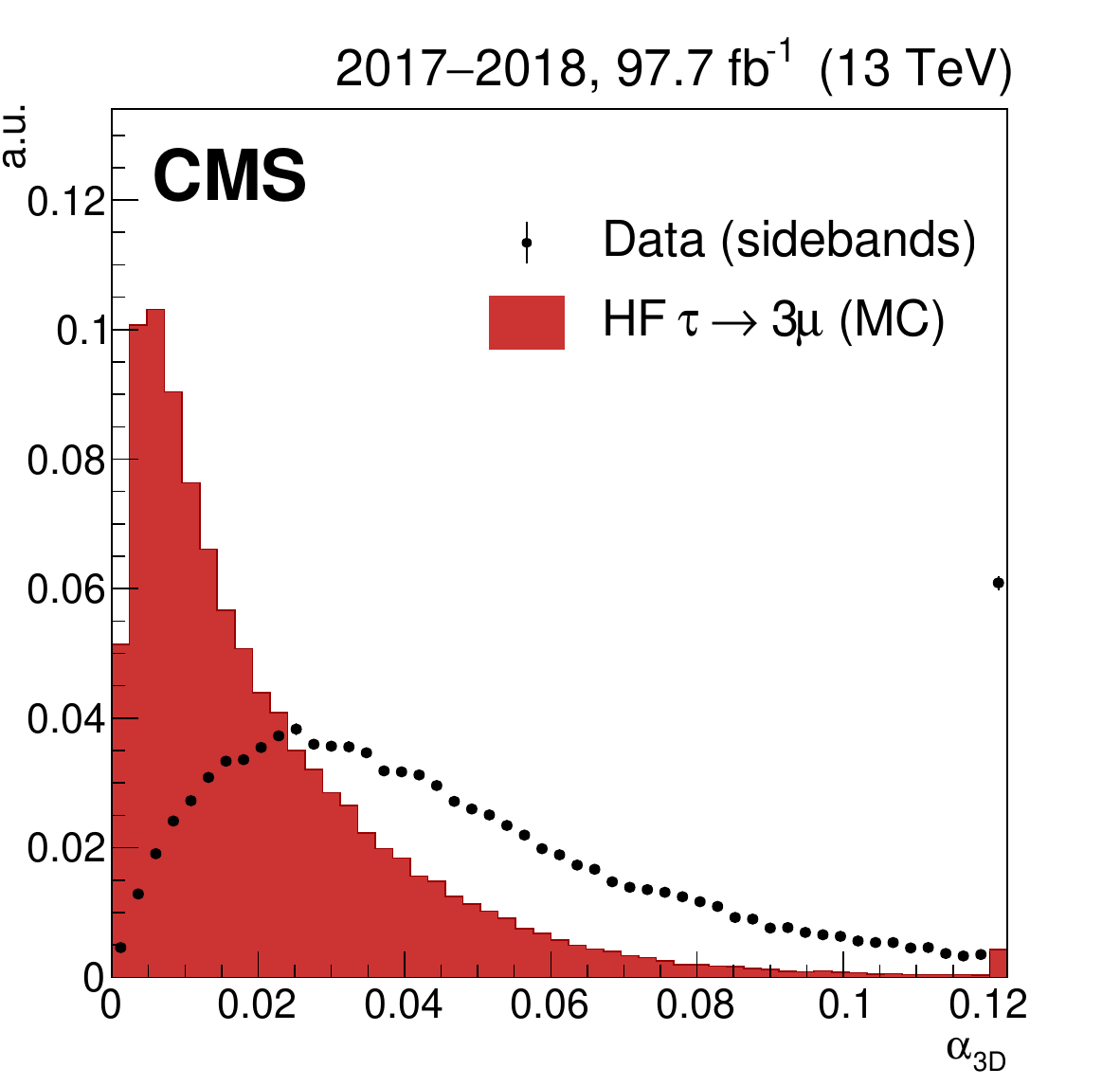}
    \includegraphics[width=2.2in]{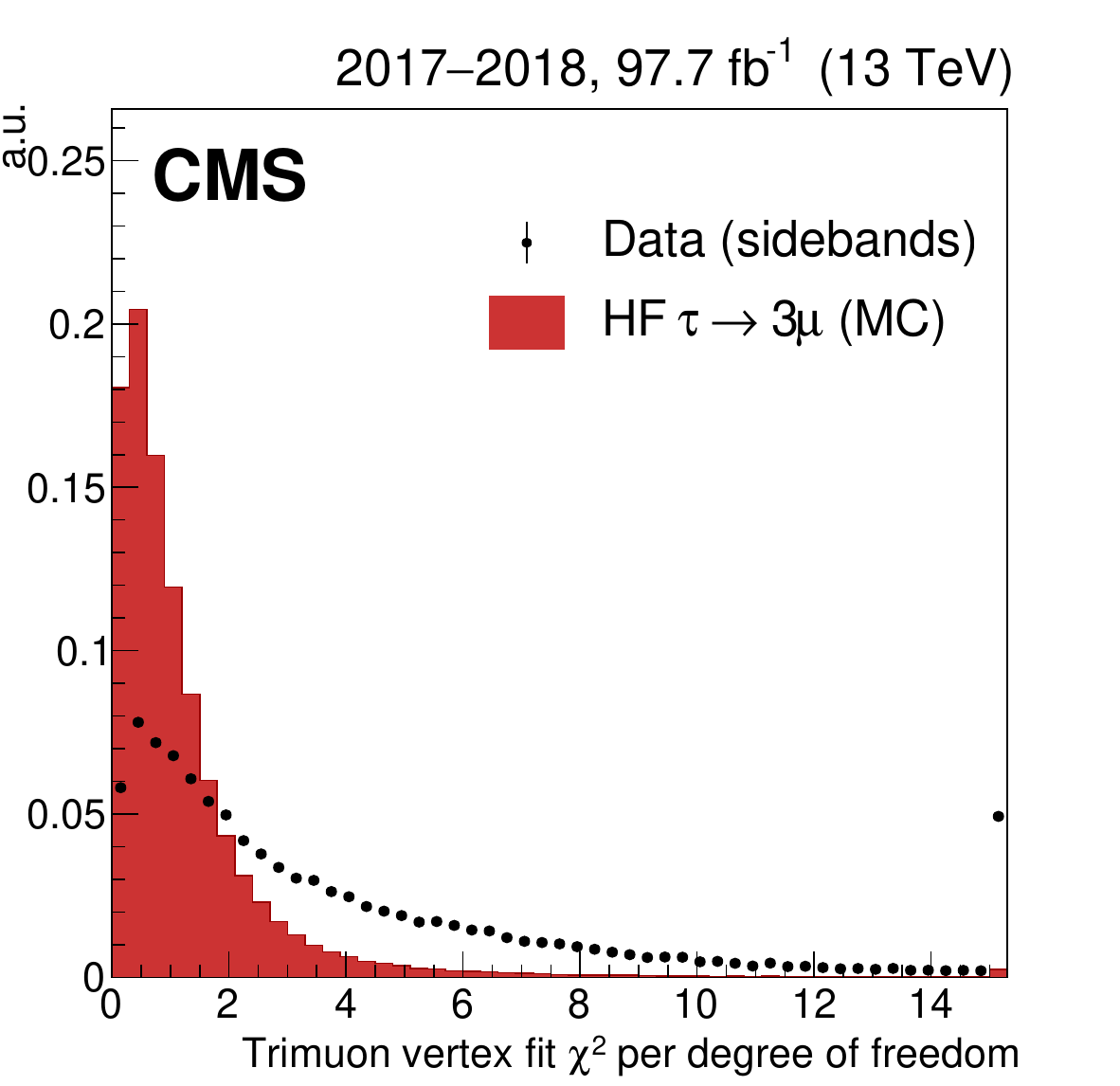}
    \includegraphics[width=2.2in]{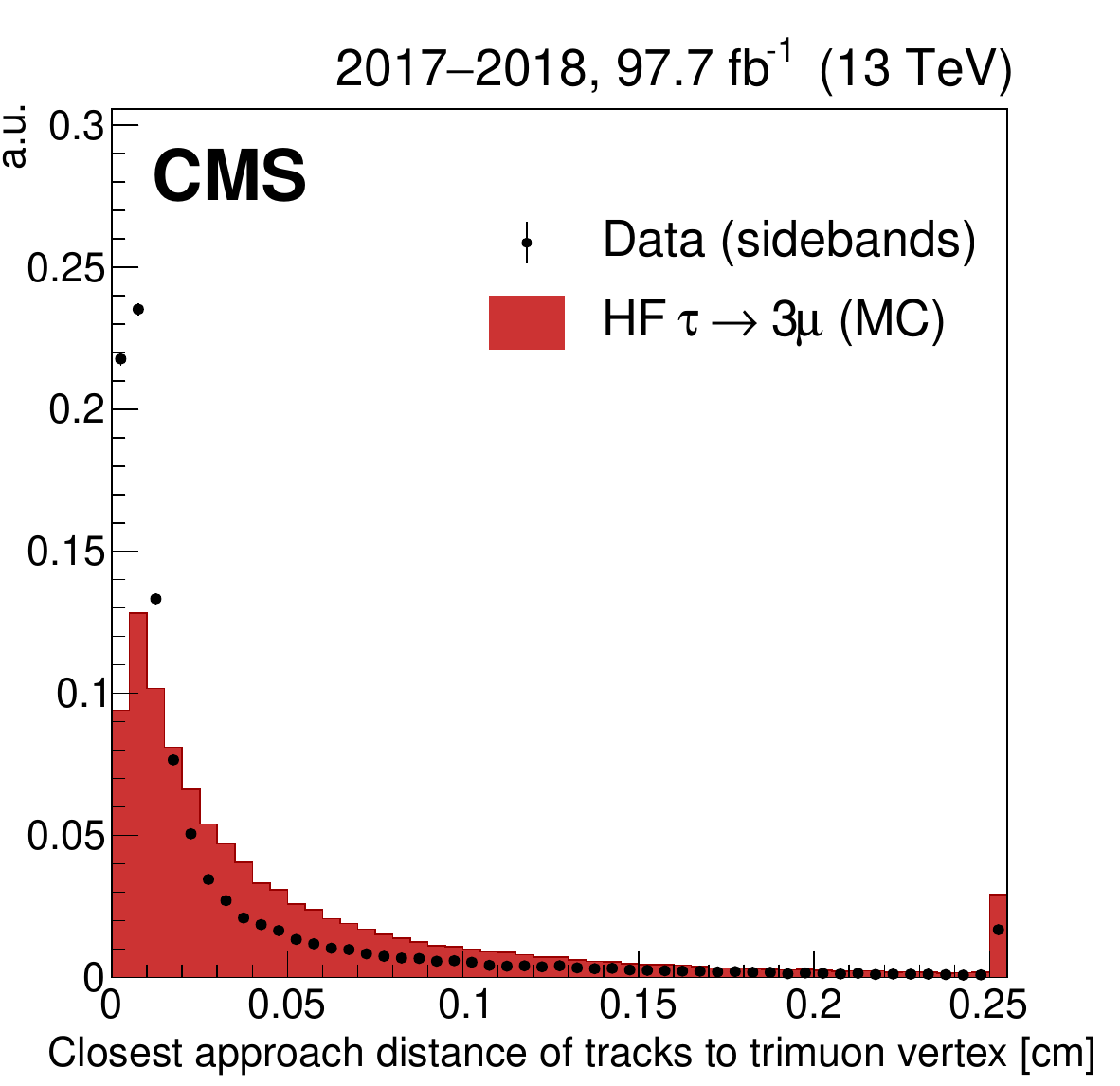}
    \includegraphics[width=2.2in]{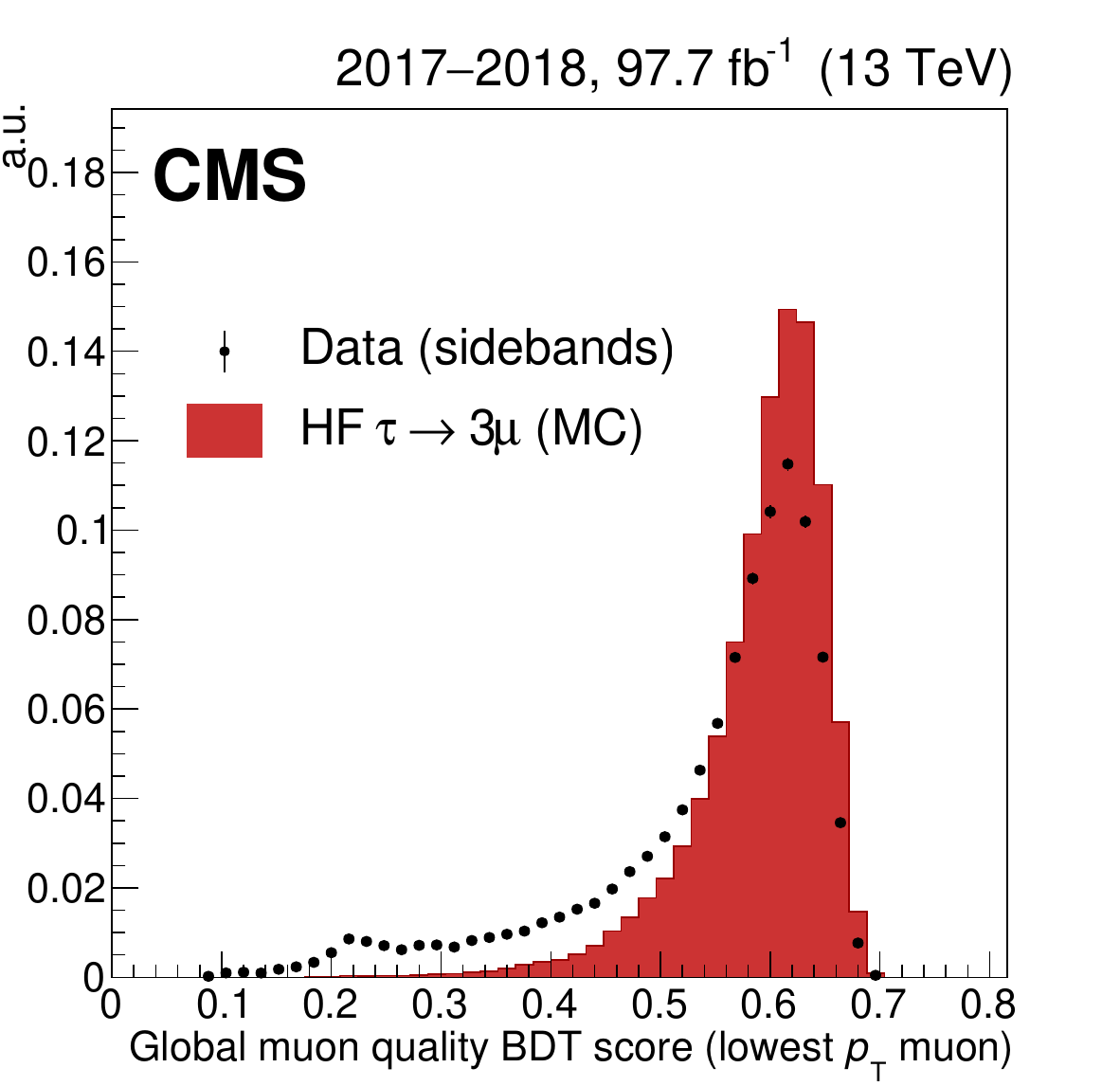}
\vspace*{8pt}
\caption{Signal and background distributions for the four observables with the highest discrimination
power used for the heavy-flavor (HF) analysis BDT training: the pointing angle $\alpha$ (upper left),
$\chi^2$ per degree of freedom of the trimuon vertex fit (upper right),
the smallest distance of closest approach to the trimuon vertex of
all the other tracks in the event with \pt $> 1~GeV$ (lower left), and 
the muon reconstruction quality BDT score of the lowest \pt muon of the three (lower right).
The signal and background distributions are obtained respectively from MC simulation
and from the mass-sideband regions in data.
All distributions are normalized to unit area.
The rightmost bins include overflow.
\protect\label{HF_inputs}}
\end{figure}

Distributions of these observables are shown in Fig.~\ref{HF_inputs} (taken from Ref.~\refcite{CMS:2023iqy}).

Based on the analysis BDT outputs, signal and background events are further divided into several sub-categories. For example, category A is divided into A1, A2, A3, and A4, in the order of decreasing BDT score (consequently decreasing signal-to-background ratio). The boundaries between the 4 sub-categories are optimized for the largest expected signal significance of A1, A2 and A3 combined, while A4 is discarded.

\begin{figure}[hbtp]
\centerline{
\includegraphics[width=2.2in]{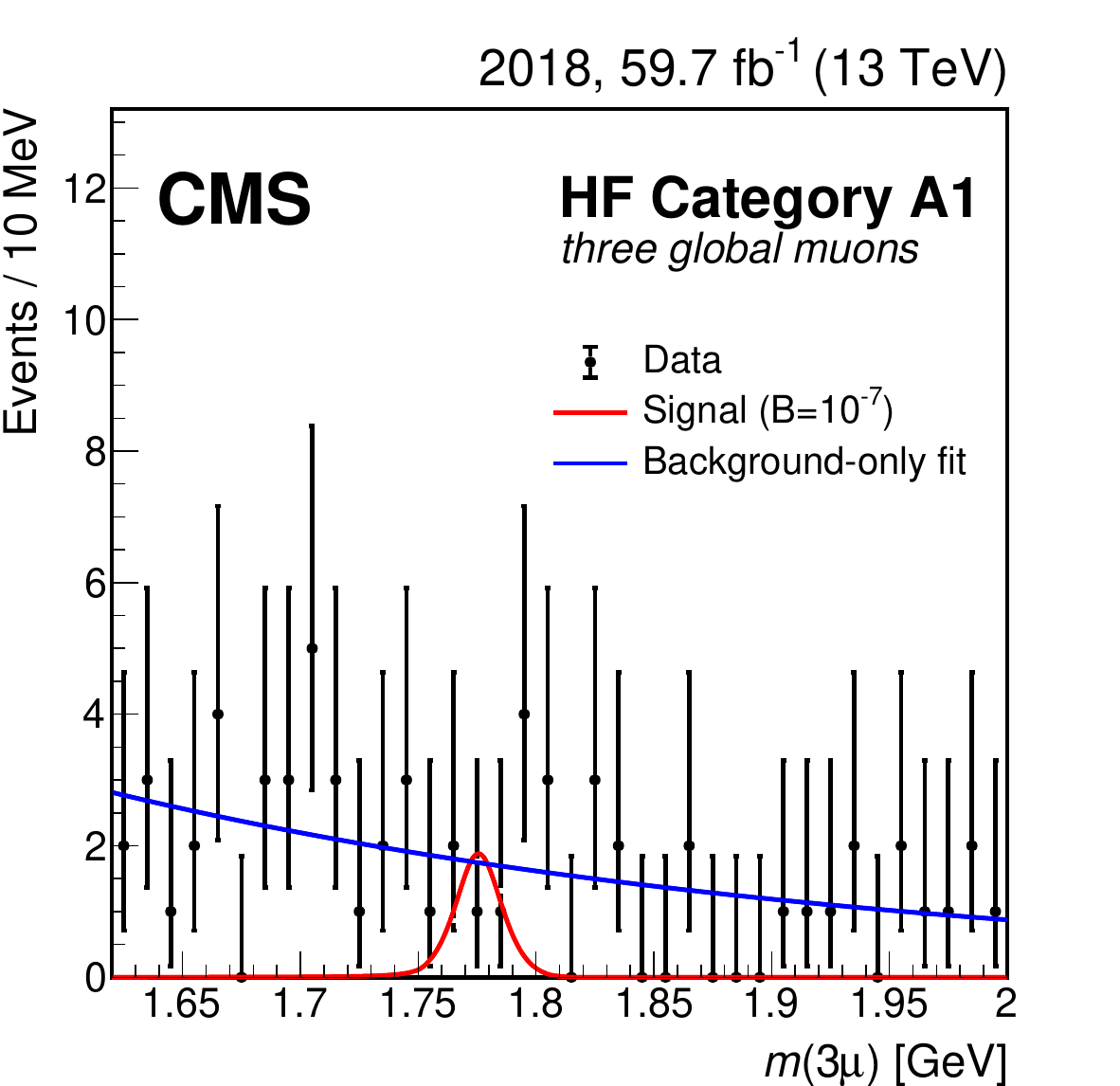}
\includegraphics[width=2.2in]{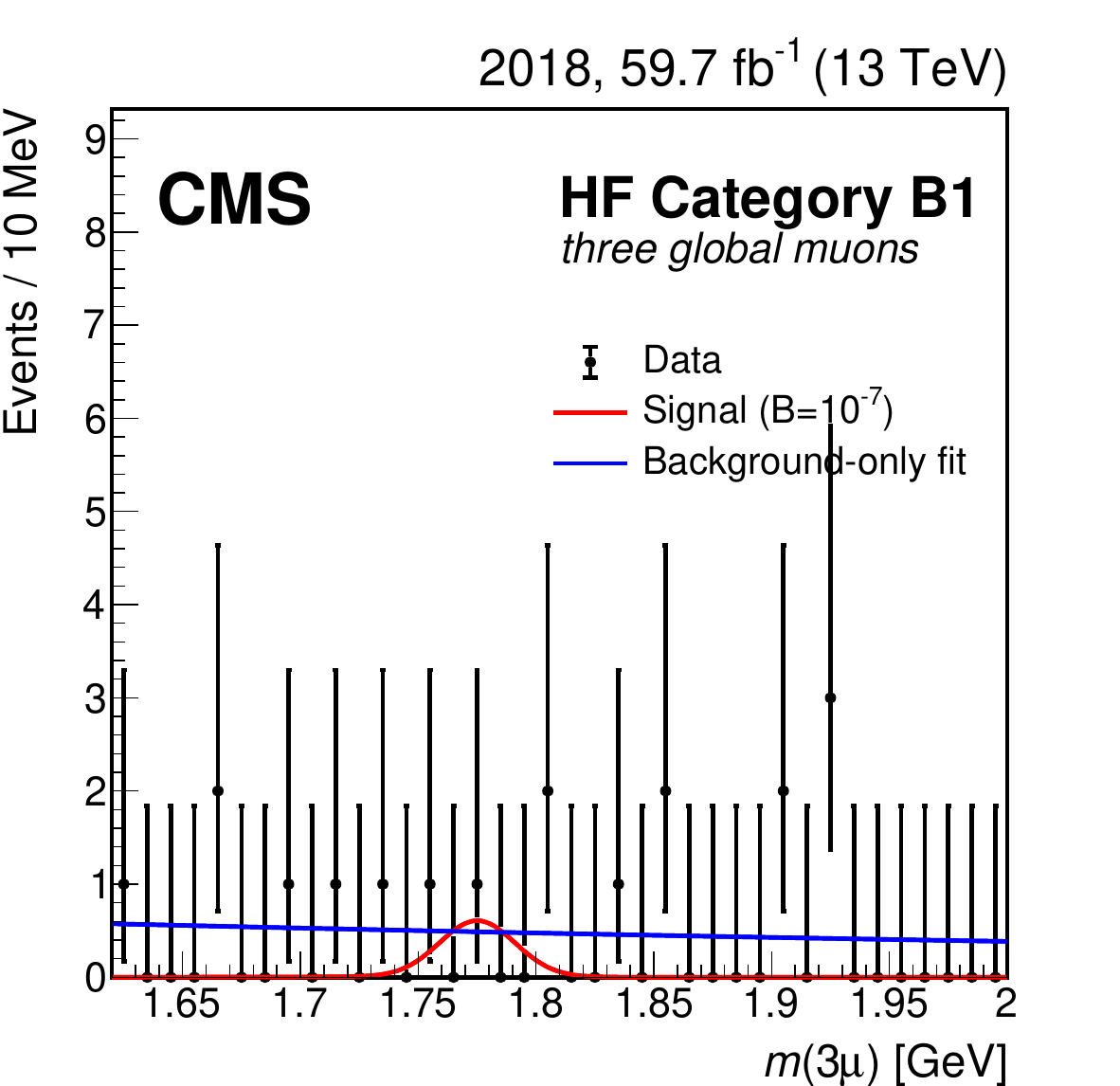}
}
\vspace*{8pt}
\caption{Representative trimuon mass distributions in the highest BDT score event categories of the heavy-flavor (HF) analysis: 2018 A1 (left) and 2018 B1 (right). Data are shown with black markers. The background-only fit and the expected
signal for $\mathcal{B}(\tau\to3\mu) = 10^{-7}$ are shown with blue and red lines, respectively.
\protect\label{HF_mass}}
\end{figure}

The final trimuon mass distributions in some of the best signal-to-background ratio categories are shown in Fig.~\ref{HF_mass} (taken from Ref.~\refcite{CMS:2023iqy}). A signal shape, obtained by fitting the simulated signal using Gaussian and Crystal Ball function, is superimposed, and $\mathcal{B}$(\ttm) = $10^{-7}$ is assumed for visibility.

In the analysis described above, the normalization channel $D_s^+ \to \phi(\mu\mu) \pi^+$ is extensively used:
\begin{itemlist}
 \item An overall signal yield scale factor (and its associated uncertainty) is obtained by comparing the $D_s^+ \to \phi(\mu\mu) \pi^+$ yields in data and simulation.
 \item All BDT input observables are validated by comparing the same distributions of $D_s^+ \to \phi(\mu\mu) \pi^+$ in data and simulation.
 \item The potential difference in BDT response to data and simulation is studied by training another BDT using simulated $D_s^+ \to \phi(\mu\mu) \pi^+$ as signal, with exactly the same input observables as in the \ttm~analysis BDT. Then the selection efficiencies for $D_s^+$ in data and simulation are compared for various BDT selection thresholds, the difference of which (about 10\%) is assigned as a systematic uncertainty.
 \item The potential differences in muon momentum scale and resolution between data and simulation are studied by fitting the $D_s^+$ peak in data and comparing its mean and width with simulation prediction in various $|\eta|$ regions.
\end{itemlist}

A simultaneous unbinned maximum likelihood fit to the trimuon mass distributions in all categories is performed, in which the \ttm~decay branching fraction is the common parameter of interest. The background in each category could be modeled using an exponential, a power law, or a second-order polynomial function. The systematic uncertainty associated with the choice of the function is treated as a discrete nuisance parameter~\cite{Dauncey:2014xga}. No significant signal is seen. The upper limit on $\mathcal{B}$(\ttm) is found to be $3.4 \times 10^{-8}$ at 90\% CL, with an expected limit of $3.6 \times 10^{-8}$.

\subsection{$\tau$ from W boson decays}

The \ttm~decays where the $\tau$ originates from a W boson have the $W\to l \nu$ characteristics that are commonly used in a W boson cross section measurement - only the $l$ is replaced by a trimuon system. These characteristics include:

\begin{itemlist}
 \item Average \pt of the trimuon is about half of the W boson mass $m(W)$;
 \item The trimuon is isolated from hadronic activities in the event;
 \item Significant $p^{miss}_T$ due to the undetected neutrino;
 \item The transverse mass, $m_T$, defend as $\sqrt{\smash[b]{2p_T^{\vphantom{a}\smash{\tau}}p^{miss}_T(1 - \cos{\Delta\phi(\vec{p_T}^{\vphantom{a}\smash{\tau}}, \vec{p}^{~miss}_T)})}}$, is consistent with $m(W)$.
\end{itemlist}

The analysis uses the same L1 triggers as those in the HF analysis. At HLT, the 3 muons must have \pt $> 7, 1, 1$ GeV, respectively, and the trimuon system must have \pt $> 15$ GeV.

The analysis strategy, similar to that in the HF analysis, is to use loose offline pre-selection criteria in order to preserve efficiencies; after that, BDT is trained to distinguish signal and background; events are categorized based on the trimuon mass resolution; and a simultaneous unbinned maximum likelihood fit to the trimuon mass distributions in all categories is performed to exact the signal.

BDT is trained using simulated signal and data sideband events. The most discriminating input observables are:

\begin{itemlist}
 \item The pointing angle $\alpha$;
 \item The isolation of the trimuon, defined as the scalar \pt sum of photons and other tracks in the vicinity of the trimuon divided by the trimuon \pt;
 \item The trimuon vertex fit quality;
 \item The trimuon \pt.
\end{itemlist}

\begin{figure}[hbtp]
    \centering
    \includegraphics[width=2.2in]{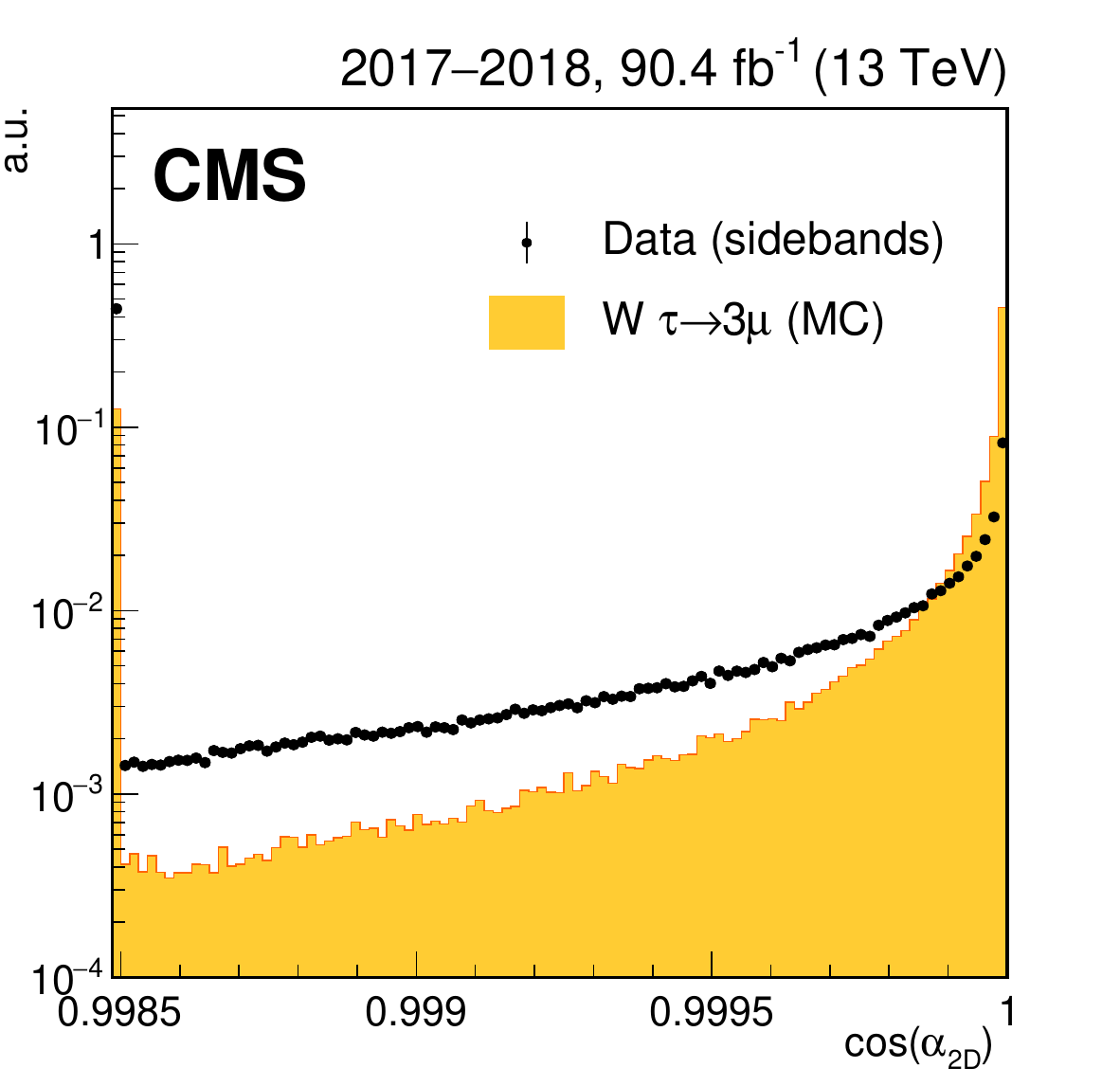}
    \includegraphics[width=2.2in]{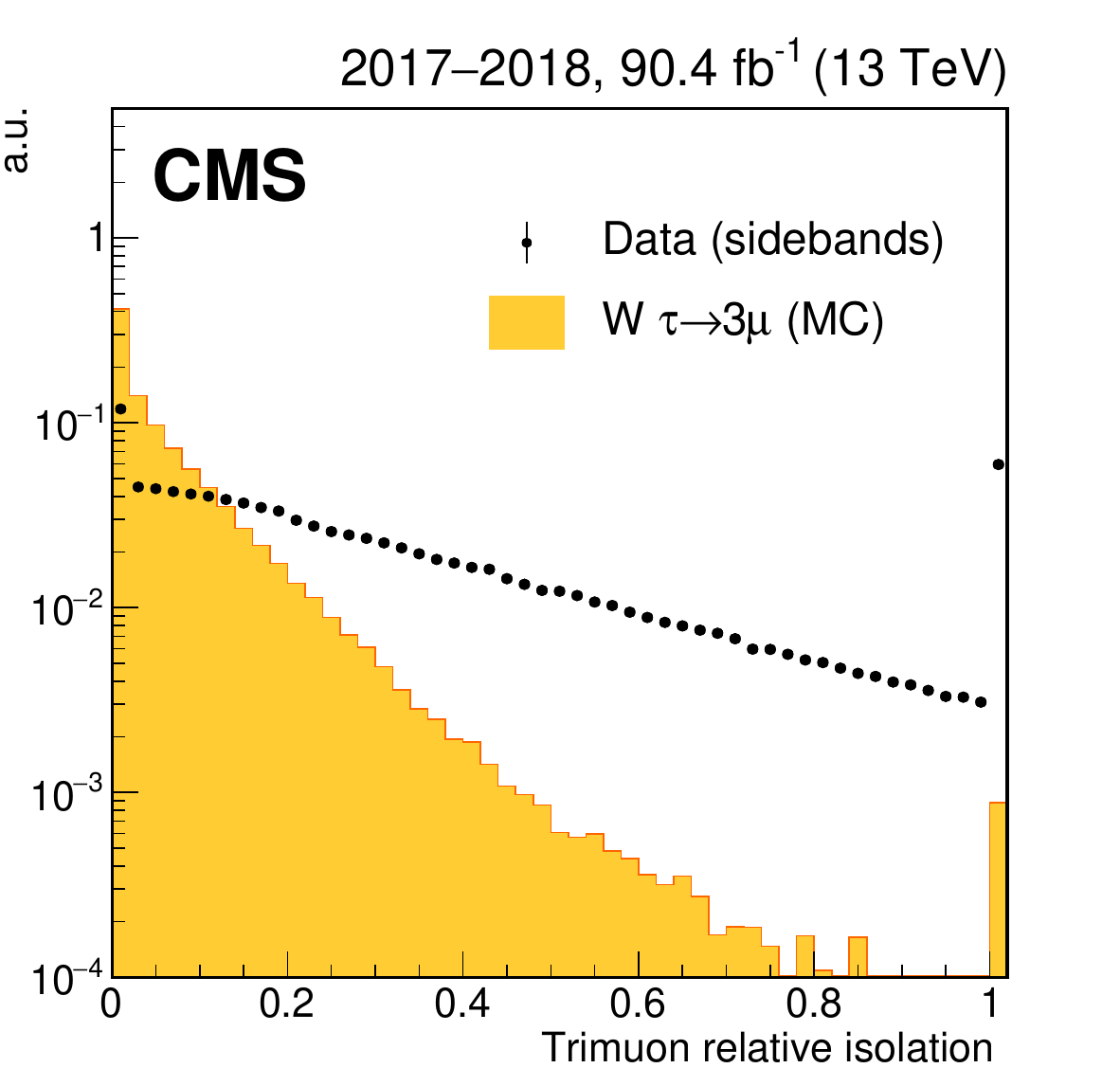}
    \includegraphics[width=2.2in]{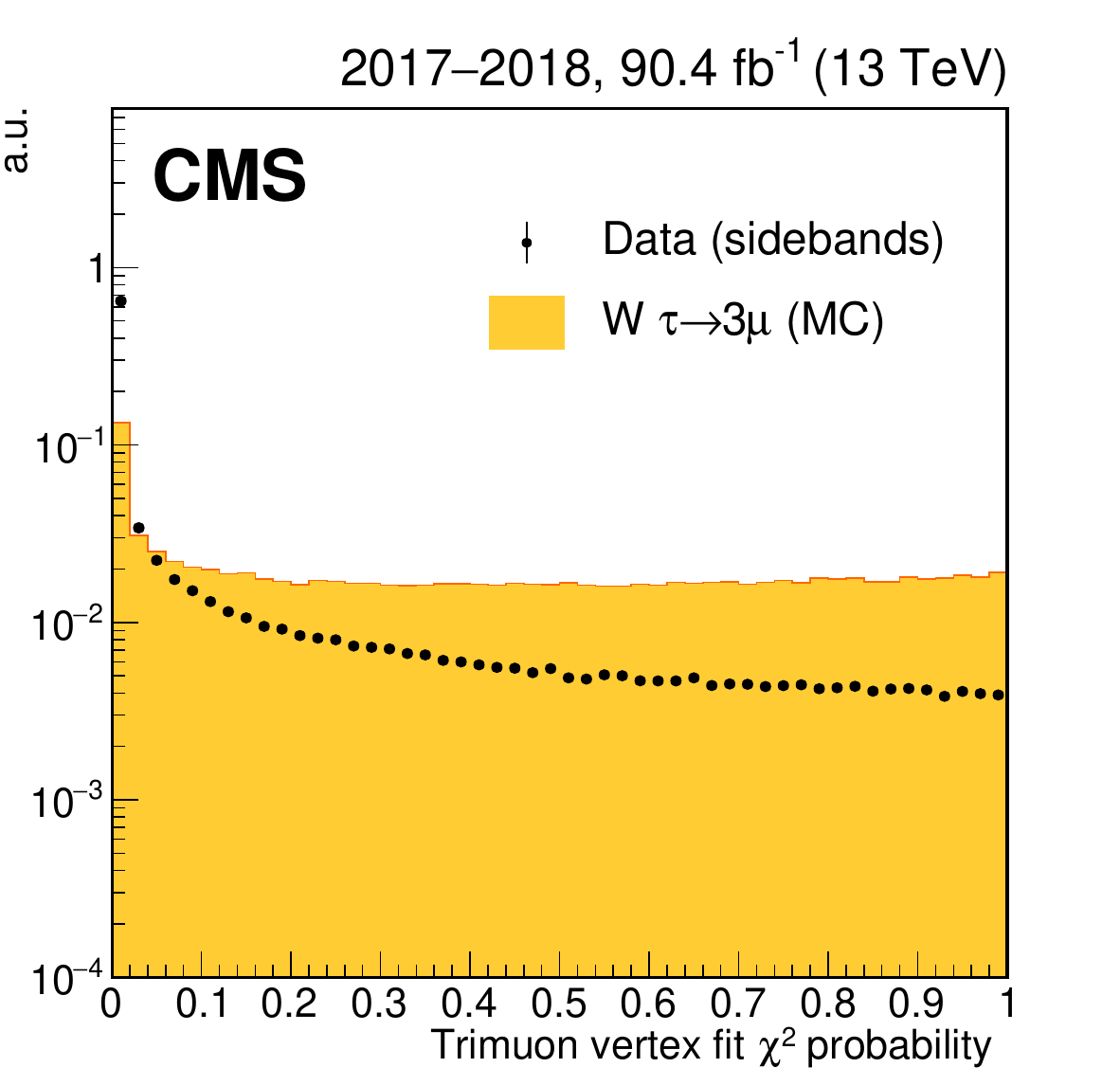}
    \includegraphics[width=2.2in]{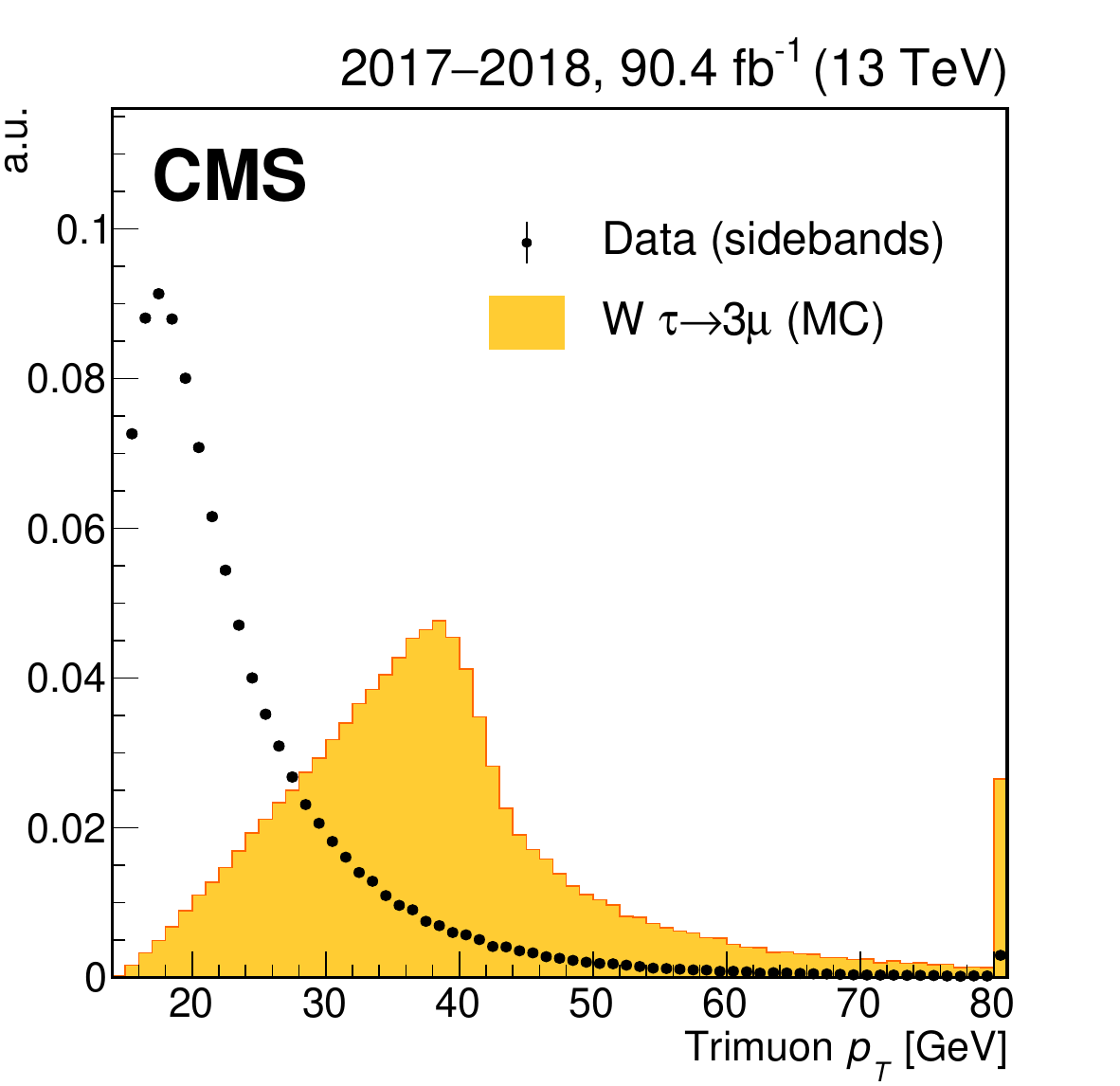}
\vspace*{8pt}
\caption{Signal and background distributions for the four observables with the highest discrimination power used for the W boson analysis BDT training:  cos($\alpha$) (upper left), relative isolation observable of the trimuon system (upper right), trimuon vertex fit $\chi^2$ probability (lower left), trimuon \pt (lower right). The signal and background distributions are obtained respectively from MC simulation and from the mass-sideband regions in data. All distributions are normalized to unit area. The leftmost (rightmost) bins include underflow (overflow).
\protect\label{W_inputs}}
\end{figure}

The distributions of these observables in simulated signal and data sideband events are shown in Fig.~\ref{W_inputs} (taken from Ref.~\refcite{CMS:2023iqy}). Other useful observables include $m_T$, the trimuon vertex displacement from the pp collision vertex, etc.

\begin{figure}[hbtp]
\centerline{
\includegraphics[width=2.in]{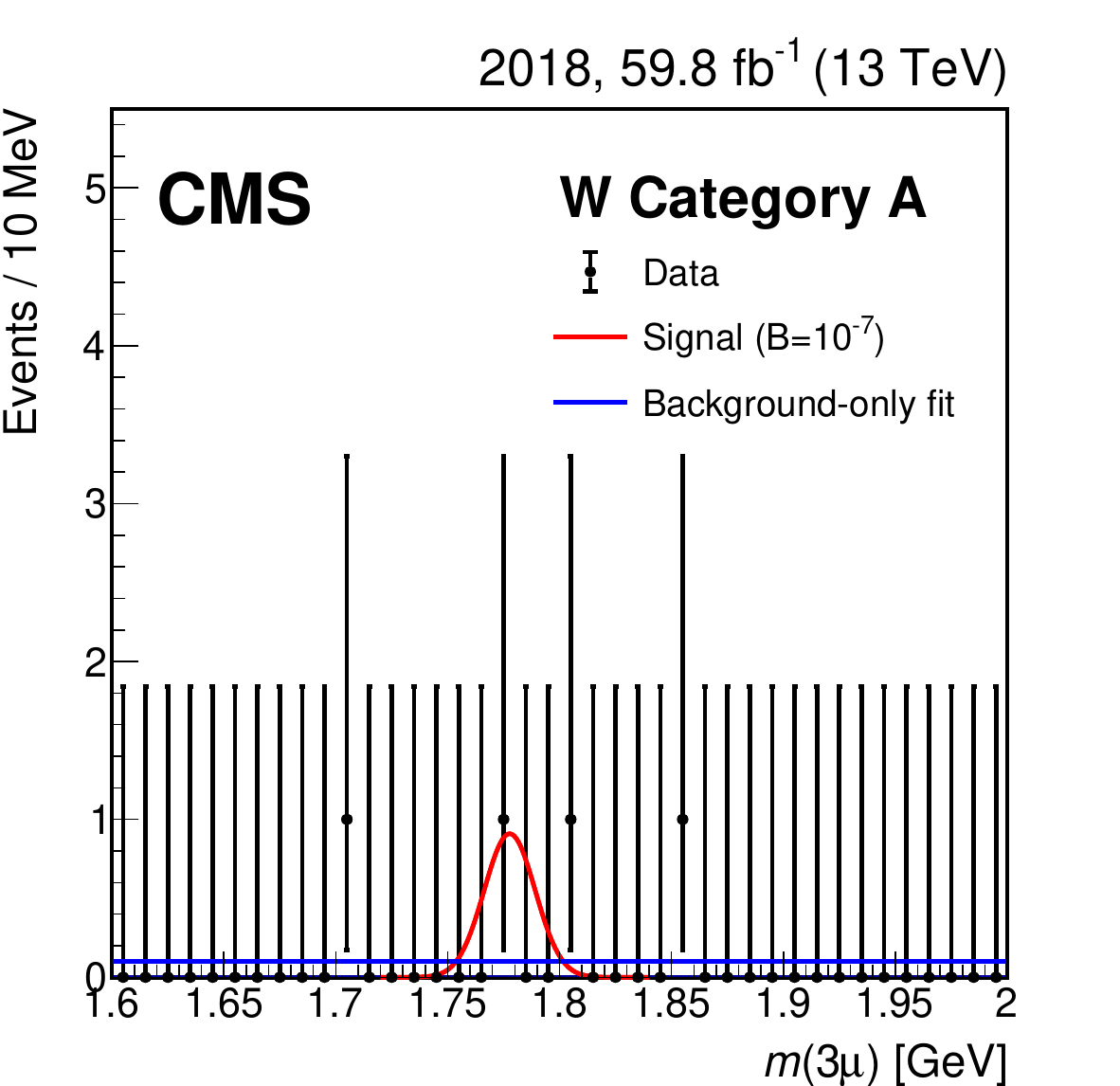}
\includegraphics[width=2.in]{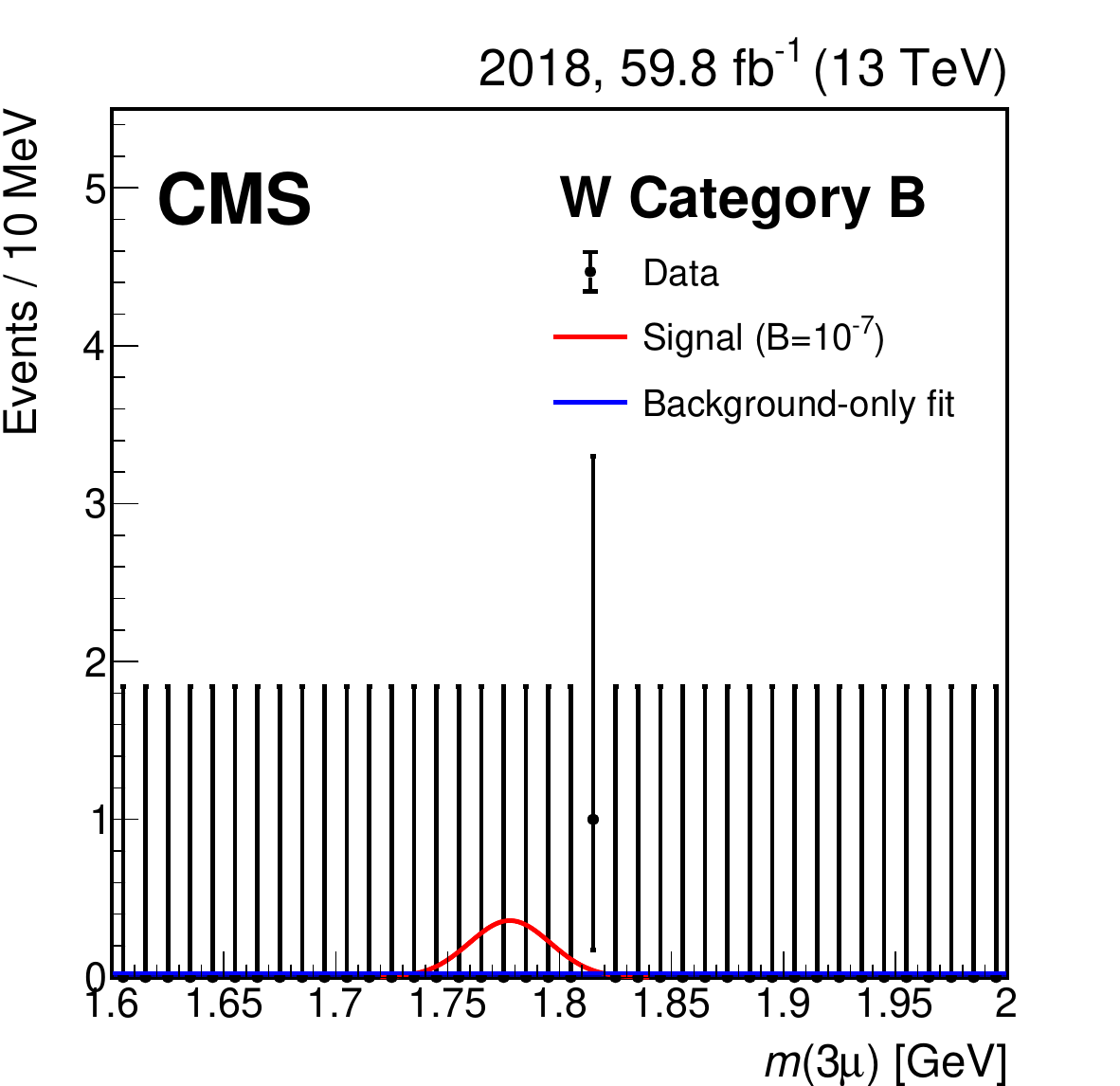}
\includegraphics[width=2.in]{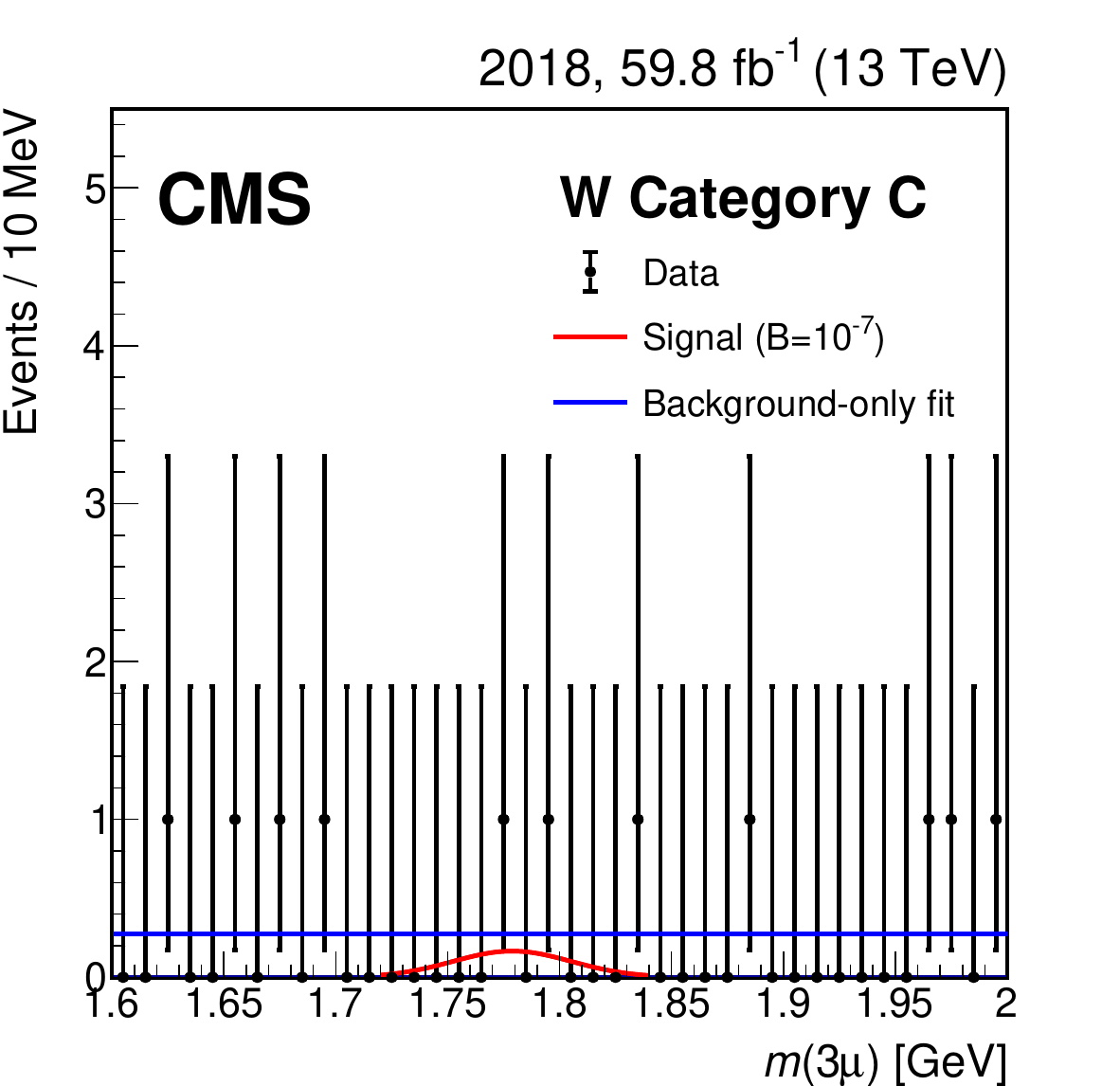}
}
\vspace*{8pt}
\caption{Trimuon mass distributions of 2018 data events in the three mass resolution categories A (left), B (middle), and C (right) of the W boson analysis. Data are shown with black markers. The background-only fit and the expected signal for $\mathcal{B}(\tau\to3\mu) = 10^{-7}$ are shown with blue and red lines, respectively.
\protect\label{W_mass}}
\end{figure}

The final trimuon mass distributions in category A, B, C (from the best mass resolution to the worst) after the BDT selections, are shown in Fig.~\ref{W_mass} (taken from Ref.~\refcite{CMS:2023iqy}). A signal shape, obtained by fitting the simulated signal events using a Gaussian function, is superimposed, and $\mathcal{B}$(\ttm) = $10^{-7}$ is assumed for visibility. The background is modelled using a flat function, given there are very few events that survive the final selections.

No significant signal is seen. The upper limit on $\mathcal{B}$(\ttm) is found to be $8.0 \times 10^{-8}$ at 90\% CL, with an expected limit of $5.6 \times 10^{-8}$.

\subsection{Combination}

The HF and the W analyses, together with a previous study based solely on the 2016 data~\cite{CMS:2020kwy}, are combined. The combination is rather straightforward, with basically two decisions to make: what to do with the data events that have passed the final selections of both HF and W analyses; and how to handle the correlations of systematic uncertainties between the two analyses. The common events are removed from the HF analysis to benefit from the higher signal-to-background ratio in the W analysis. Systematic uncertainties are assumed to be uncorrelated between the two analyses, taking into account that the analysis sensitivities are limited by statistical uncertainties. In fact, the results would change by no more than a few percent if all systematic uncertainties were assumed to be zero.

\begin{figure}[hbtp]
\centerline{
\includegraphics[width=3.0in]{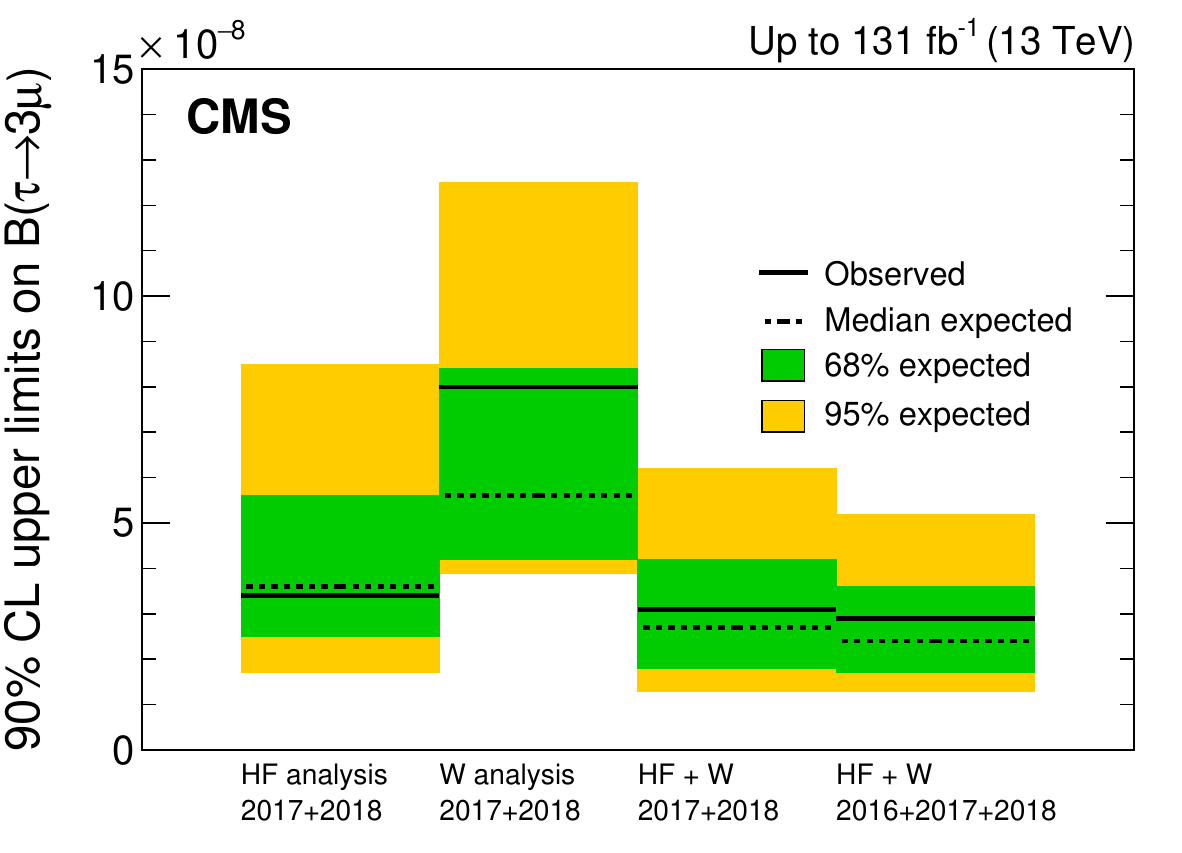}
}
\vspace*{8pt}
\caption{Observed and expected upper limits on $\mathcal{B}(\tau\to3\mu)$ at 90\% CL, from the heavy-flavor (HF) analysis, the W boson analysis, the combination of the two analyses, as well as their combination with the previously published result using 2016 data.
\protect\label{combined}}
\end{figure}

The observed upper limit on $\mathcal{B}(\tau\to 3\mu)$ is determined to be $2.9\times10^{-8}$ at 90\% CL, with an expected limit of $2.4\times10^{-8}$. Figure~\ref{combined} (taken from Ref.~\refcite{CMS:2023iqy}) shows the results of the HF analysis, the W analysis, the combination of the two, as well as the final CMS Run 2 result.

\subsection{Discussion}

In the HF analysis, the overall signal acceptance times efficiency is about 0.03\% (depending on the final BDT selection thresholds). The main challenges of the analysis are the L1 trigger acceptance and the background from low \pt misidentified muons. These two points are correlated: looser trigger thresholds could collect more signal candidate events, but these events have on average lower \pt distributions and thus lower signal-to-background ratio.

The W analysis, on the contrary, has almost zero background in its most sensitive event categories. Characteristics of W boson decays, such as high \pt leptons, large $p^{miss}_T$, and lepton isolation, offer powerful handles to suppress backgrounds. The search sensitivity of the W analysis therefore will grow faster than $\sqrt{\mathcal{L}}$, $\mathcal{L}$ being the integrated luminosity, and soon bypass that of the HF analysis. Moreover, it gives hope that other $\tau$ LFV decay mode searches might be feasible using W-sourced  $\tau$ leptons at the LHC.

Analysis-technique wise, the W analysis is less complex, without many event categories. A large amount of work was spent on the muon trigger and selection efficiency corrections, taking the strategy of measuring per-muon efficiencies in control samples of $Z\to2\mu$ or $J/\Psi\to2\mu$. It could potentially be  simplified by extracting the efficiencies of the trimuon as a whole with a normalization channel. For example, in $Z\to4\mu$ events ($Z\to2\mu$ and a pair of muons from final state radiation), 3 muons are relatively collimated, while the Z peak guarantees a pure $Z\to4\mu$ control sample. Furthermore, if the fourth muon is removed by hand, in other words replaced by an undetectable neutrino, the event becomes similar to \ttm~from a W boson decay, and could be helpful to validate the BDT analysis. 

The future High-Luminosity LHC (HL-LHC) has the goal of recording pp collision data corresponding to an integrated luminosity of  $3~ab^{-1}$, about 25 times that of the Run 2 dataset. Since the present analyses are limited by statistical uncertainties, the 25 times larger dataset will by construction result in a factor of 5 better search sensitivity (the W analysis gains even more, as mentioned above). 
The \ttm~search would also benefit from the upcoming   Phase-2 detector upgrade~\cite{Contardo:2015bmq}. 
All in all, the CMS \ttm~search is going to approach a sensitivity of $10^{-9}$ with the full HL-LHC data.

Though the HL-LHC projection looks promising, we shall not forget that the leading player on $\tau$ LFV decay searches for the coming decade is the Belle II experiment. With the target integrated luminosity of $50~ab^{-1}$, Belle II is expected to set sub-$10^{-9}$ upper limits on branching fractions of \ttm~and a few other $\tau$ decay modes. The LHC experiments are not limited by the amount of $\tau$ leptons, but rather by the quantity of background. A breakthrough in analysis techniques is required for the LHC experiments to remain competitive.

\section{Search for $H \to \mu \tau$ and $H \to e \tau$ decays}
\label{sec:higgs}

The SM predicts that the Higgs field couples to fermions (quarks and leptons) through a Yukawa interaction, the coupling strength of which is proportional to the mass of that fermion, making it the only interaction that does not respect lepton universality. Lepton flavor is still conserved, unless the off-diagonal elements of the Yukawa interaction matrix ($Y_{\mu \tau}, Y_{e \tau}, Y_{e \mu}$) are nonzero, giving rise to LFV Higgs boson decays, such as $H \to \mu \tau$ or $H \to e \tau$.
The CMS searches for these decays using Run 2 data is detailed here.

\begin{figure}[hbtp]
\centerline{
\includegraphics[width=2.6in]{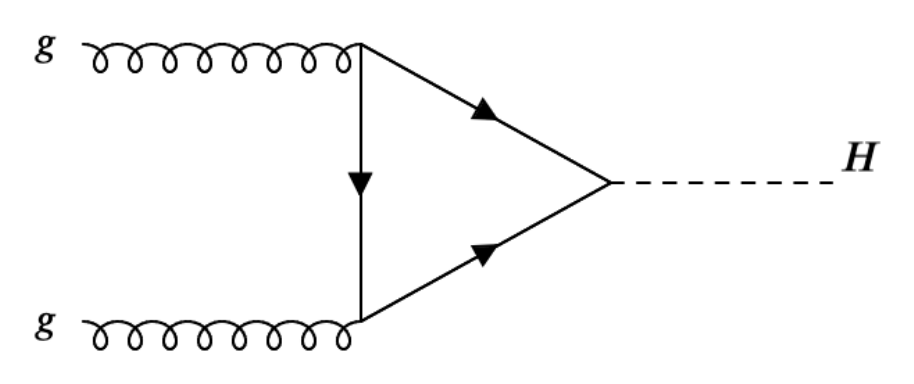}
\includegraphics[width=1.8in]{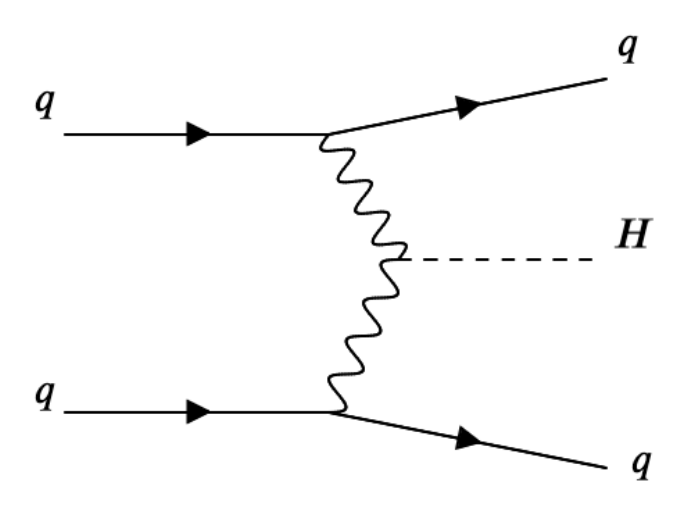}
}
\vspace*{8pt}
\caption{Feynman diagrams of the two main Higgs boson production modes: gluon fusion (left) and vector boson fusion (right).
\protect\label{higgs_diagrams}}
\end{figure}

The analysis targets the two main Higgs boson production modes, gluon fusion (ggH) and vector boson fusion (VBF), the Feynman diagrams of which are shown in Figure~\ref{higgs_diagrams}.  Contributions from other Higgs boson production modes are  found to be negligible. The decay final state depends on the $\tau$ lepton decays, i.e. $\tau$ to hadrons, electrons or muons, which are respectively denoted as
$\tau_h$, $\tau_e$ and $\tau_\mu$.

In the $H \to \mu \tau$ search, $\tau_h$ and $\tau_e$ are explored, but not $\tau_\mu$ due to large Drell-Yan background. For the same reason, in the $H \to e \tau$ search, only $\tau_h$ and $\tau_\mu$ are used. The trigger and offline selections require two high \pt and isolated leptons (electron, muon, or $\tau$ lepton, depending on the final states). The muon or electron must have the opposite charge as the $\tau$. Details of event selection thresholds are summarized in Table~\ref{mutau_selection} and Table~\ref{etau_selection}. ($I^{l}_{\text{rel}}$ in the tables is the relative isolation observable of $\mu$ or e.)

\begin{table}[htbp]
\tbl{Event selection criteria for the $H \to \mu \tau$ channels.}
{\begin{tabular}{ccccc}
Variable                        &   $\mu \tau_h$             &  $\mu \tau_e$                              \\[0.5ex]
\hline \\ [-2.0ex]
$p_T^{e}$                     &   -                &  $>$13~GeV                         \\[0.5ex]
$p_T^{\mu}$                    &   $>$26~GeV          &  $>$24~GeV                         \\[0.5ex]
$p_T^{\tau_h}$                   &   $>$30~GeV          &  -                               \\[1.0ex]
$|{\eta}|^{e}$              &   -                &  $<$2.5                            \\[0.5ex]
$|{\eta}|^{\mu}$             &   $<$2.1             &  $<$2.4                            \\[0.5ex]
$|{\eta}|^{\tau_h}$            &   $<$2.3             &  -                               \\[1.0ex]
$I^{e}_{\text{rel}}$          &   -                &  $<$0.1                            \\[0.5ex]
$I^{\mu}_{\text{rel}}$         &   $<$0.15            &  $<$0.15                           \\[1.0ex]
\multirow{2}{*}{\begin{tabular}{c}Trigger \\ requirement \end{tabular}}
                                & $p_T^{\mu}>24$~GeV (all years) &
\multirow{2}{*}{\begin{tabular}{c}$p_T^{e}>12$~GeV \\ $p_T^{\mu}>23$~GeV \end{tabular}}  \\[0.5ex]
                                &    $p_T^{\mu}>27$~GeV (2017)   &                         \\[0.5ex]
\end{tabular}
\label{mutau_selection}}
\end{table}

\begin{table*}[htbp]
\tbl{Event selection criteria for the $H \to e \tau$ channels.}
{\begin{tabular}{ccccc}
Variable                        &          $e\tau_h$                     &  $e \tau_\mu$                           \\[0.5ex]
\hline \\ [-2.0ex]
$p_T^{e}$                   &          $>27$~GeV               &  $>24$~GeV                    \\[0.5ex]
$p_T^{\mu}$                  &          -                     &  $>10$~GeV                    \\[0.5ex]
$p_T^{\tau_h}$                 &          $>30$~GeV               &  -                          \\[1.0ex]
$|{\eta}|^{e}$            &          $<2.1$                  &  $<2.5$                       \\[0.5ex]
$|{\eta}|^{\mu}$           &          -                     &  $<2.4$                       \\[0.5ex]
$|{\eta}|^{\tau_h}$          &          $<2.3$                  &  -                          \\[1.0ex]
$I^{e}_{\text{rel}}$        &          $<0.15$                 &  $<0.1$                     \\[0.5ex]
$I^{\mu}_{\text{rel}}$       &          -                     &  $<0.15$                      \\[1.0ex]
\multirow{4}{*}{\begin{tabular}{c}Trigger \\ requirement \end{tabular}}
                                &   $p_T^{e}>25$~GeV (2016)        &
\multirow{4}{*}{\begin{tabular}{c}$p_T^{e}>23$~GeV \\ $p_T^{\mu}>8$~GeV \end{tabular}}           \\[0.5ex]
                                &   $p_T^{e}>27$~GeV (2017)        &                            \\[0.5ex]
                                &   $p_T^{e}>32$~GeV (2018)        &                            \\[0.5ex]
                                & $p_T^{e}>24$~GeV and $p_T^{\tau_h}>30$~GeV (2017, 2018) &      \\[0.5ex]
\end{tabular}
\label{etau_selection}}
\end{table*}

Selected events are categorized based on the number of jets:
\begin{itemlist}
 \item 0 jet: targeting the ggH production;
 \item 1 jet: targeting the ggH with an initial-state-radiation jet;
 \item 2 jets and $m(jj) < 500$~GeV: targeting the ggH production with additional jets;
 \item 2 jets and $m(jj) > 500$~GeV: targeting the VBF production.
\end{itemlist}

The dominant background is $Z \to \tau\tau$, which is estimated using embedded samples made as described in the rest of this paragraph. First, $Z \to \mu \mu$ events are selected from data, and the two muons from the Z boson decay are removed. Then, the $\tau$ objects are taken from simulated $Z \to \tau\tau$ events, and merged with the muon-removed $Z \to \mu \mu$ data events, with the kinematics of the $\tau$ leptons matching those of the original muons. The advantages of this method include a better description of jets, pileup, detector noise, and resolution effects. 

\begin{figure}[hbtp]
    \centering
    \includegraphics[width=2.2in]{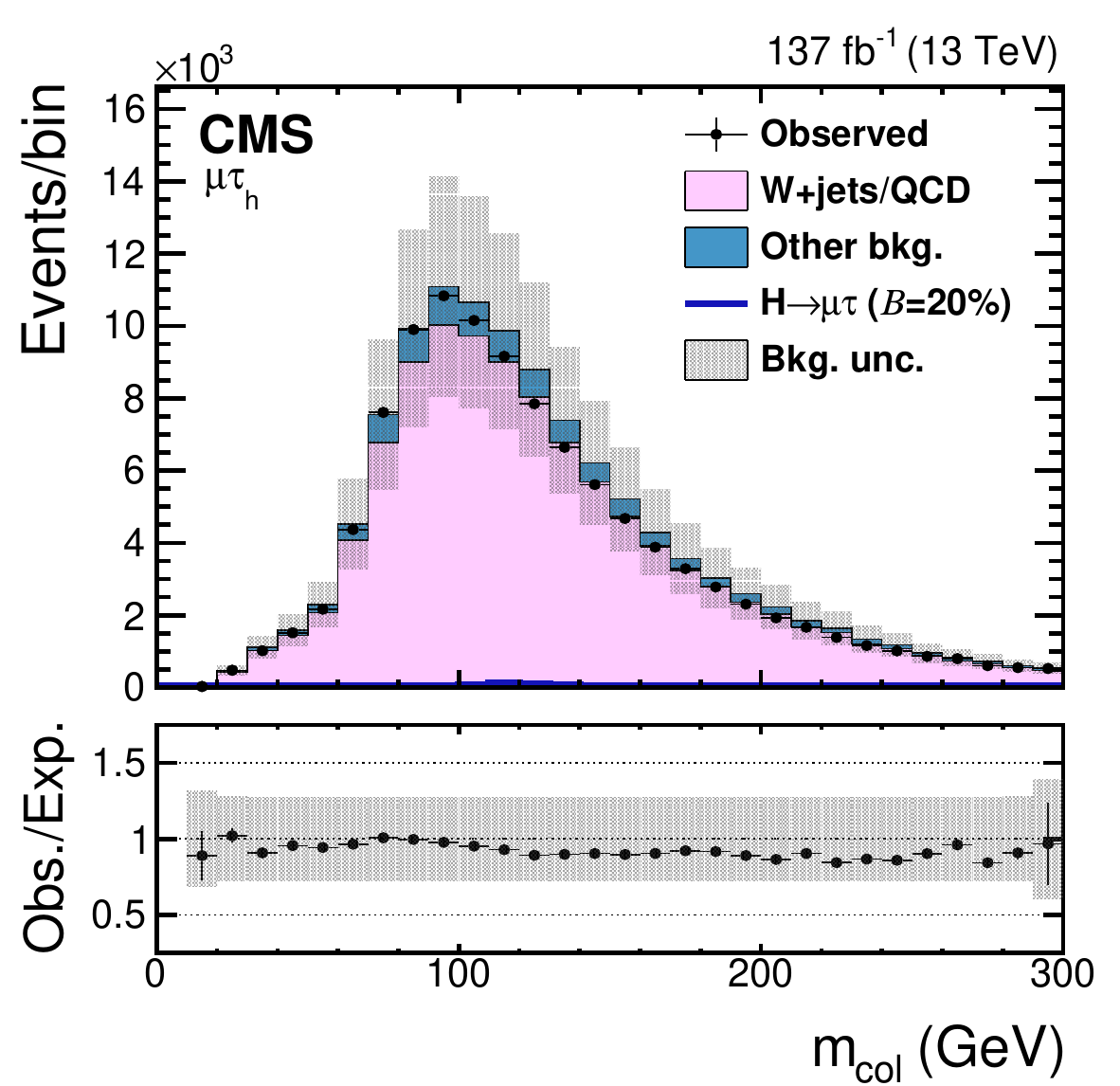}
    \includegraphics[width=2.2in]{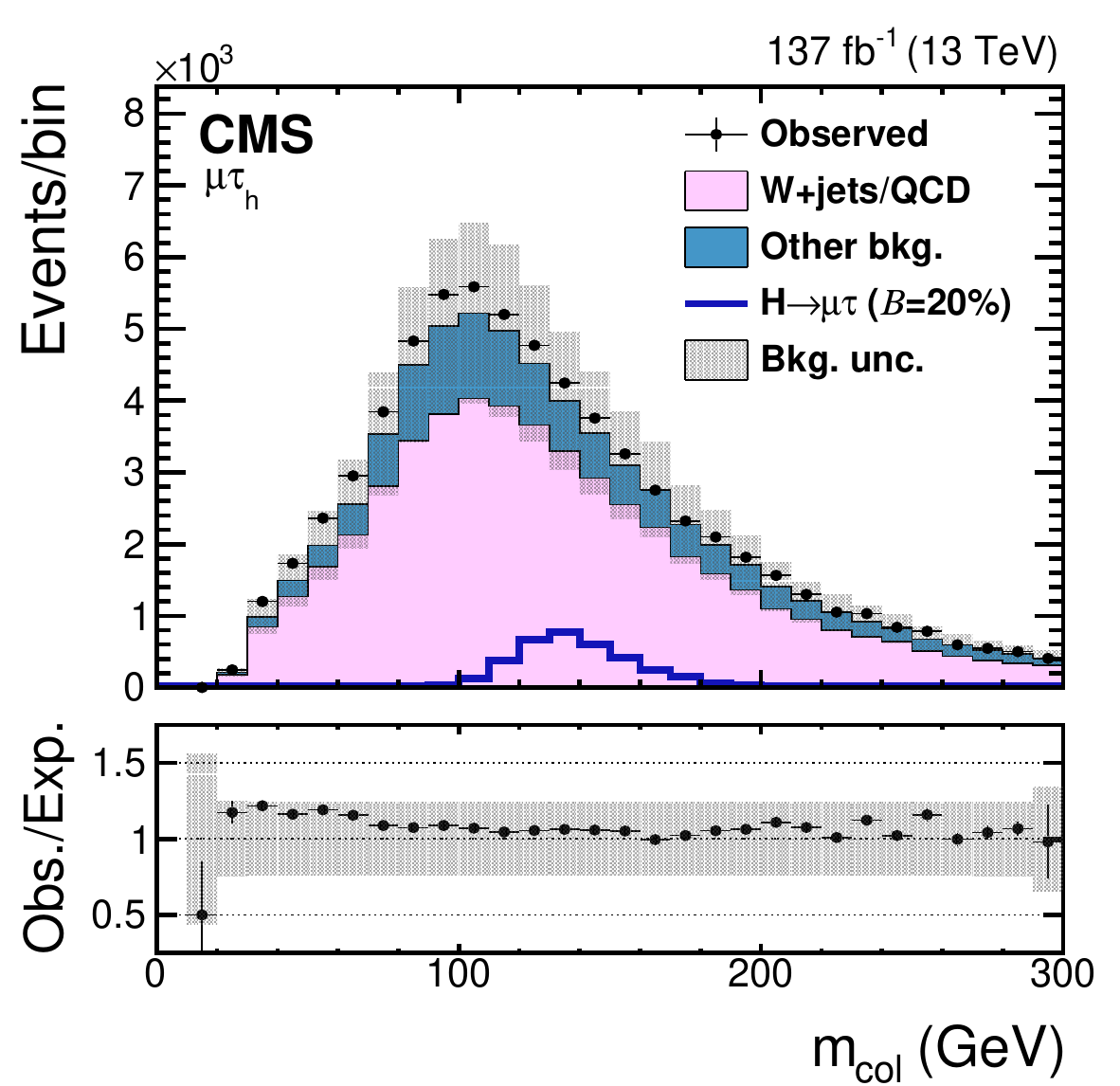}
    \includegraphics[width=2.2in]{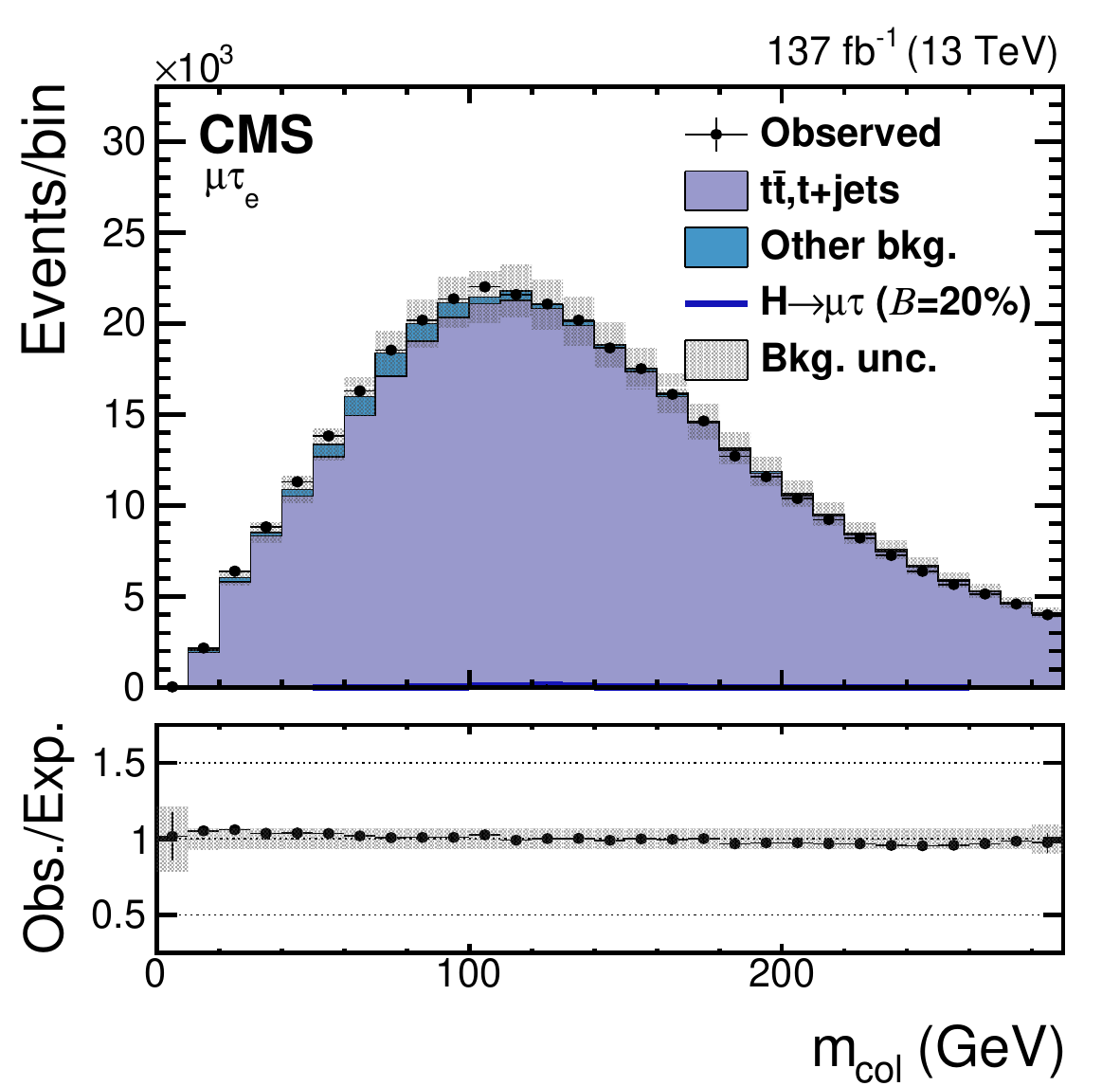}
\vspace*{8pt}
\caption{The $m_{col}$ distribution in VR with same electric charge for both leptons (top left), W+jets VR (top right), and $t\bar{t}$ VR (lower). In each distribution, the VR's dominant background is shown, and all the other backgrounds are grouped into ``Other bkg.''. A $\mathcal{B}(H \to \mu \tau)$=20\% is assumed for the signal. The lower panel in each plot shows the ratio of data and estimated background. The uncertainty band corresponds to the background uncertainty in which the post-fit statistical and systematic uncertainties are added in quadrature.
\protect\label{higgs_vr}}
\end{figure}

In addition, an important background to the $\mu\tau_h$ and $e\tau_h$ final states is from misidentified muon or electron (W+jets or QCD multi-jets). This is estimated using a “fake factor” method. Fake factors, defined as the probabilities for jets to be misidentified as muons or electrons, are measured in a control sample of Z+jets.
In the $\mu\tau_e$ and $e\tau_\mu$ final states, top quark, di-boson backgrounds are taken from simulation.
Events with b-tagged jets~\cite{CMS:2017wtu} (\pt $> 20$~GeV, $|\eta| < 2.4$) or with additional leptons or hadronic $\tau$ leptons and more than 2 jets are vetoed.
Background estimations are validated in different orthogonal validation regions (VR), as shown in Fig.~\ref{higgs_vr} (taken from Ref.~\refcite{CMS:2021rsq}). The collinear mass $m_{col}$ in the figures is an observable to estimate the invariant mass of a $\tau$ and another object, assuming the neutrino(s) from the $\tau$ decay has the momentum direction collinear with that of the other visible decay products of the $\tau$. Details could be found in Ref.~\refcite{CMS:2021rsq}.

\begin{figure}[hbtp]
    \centering
    \includegraphics[width=2.2in]{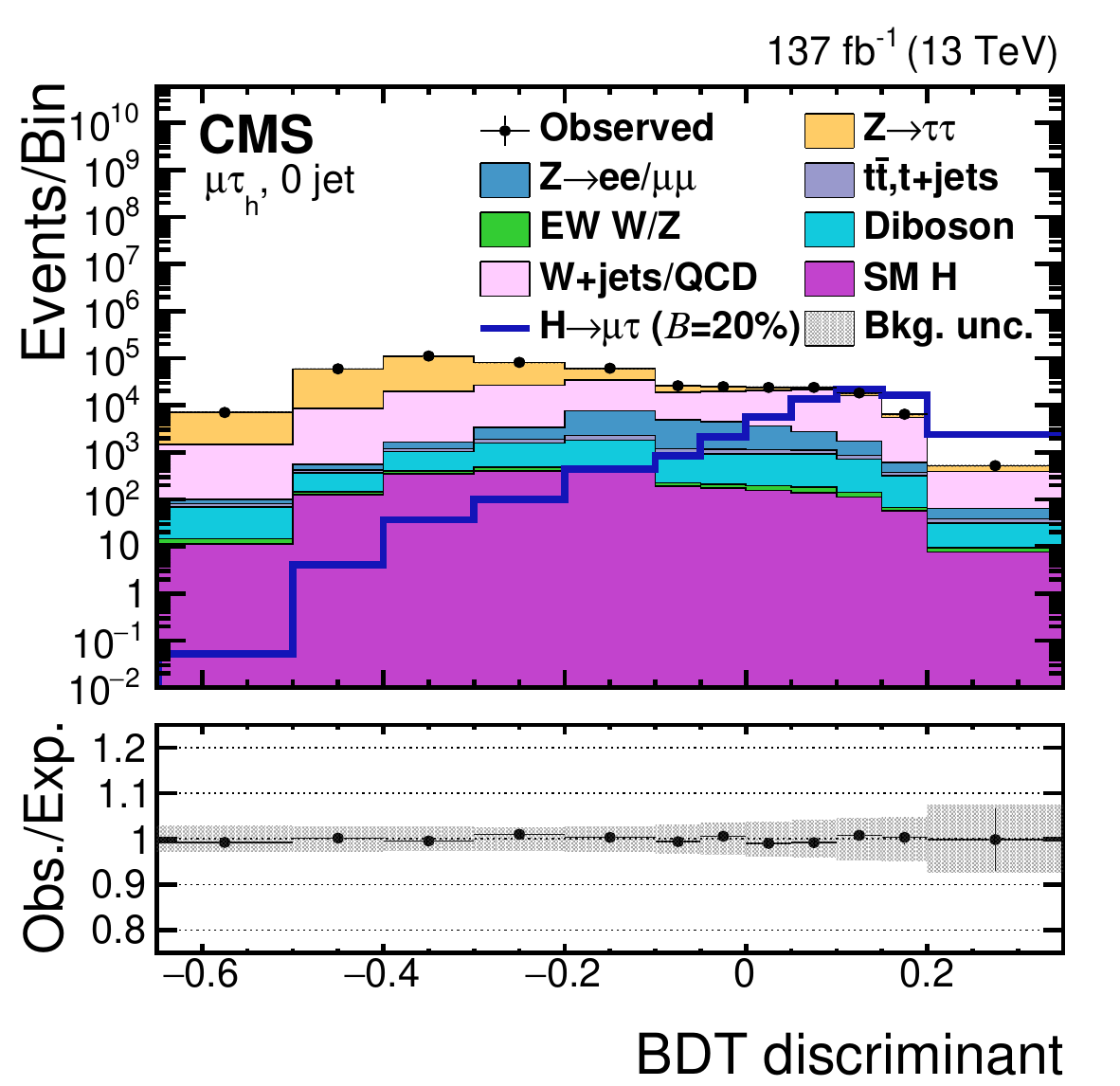}
    \includegraphics[width=2.2in]{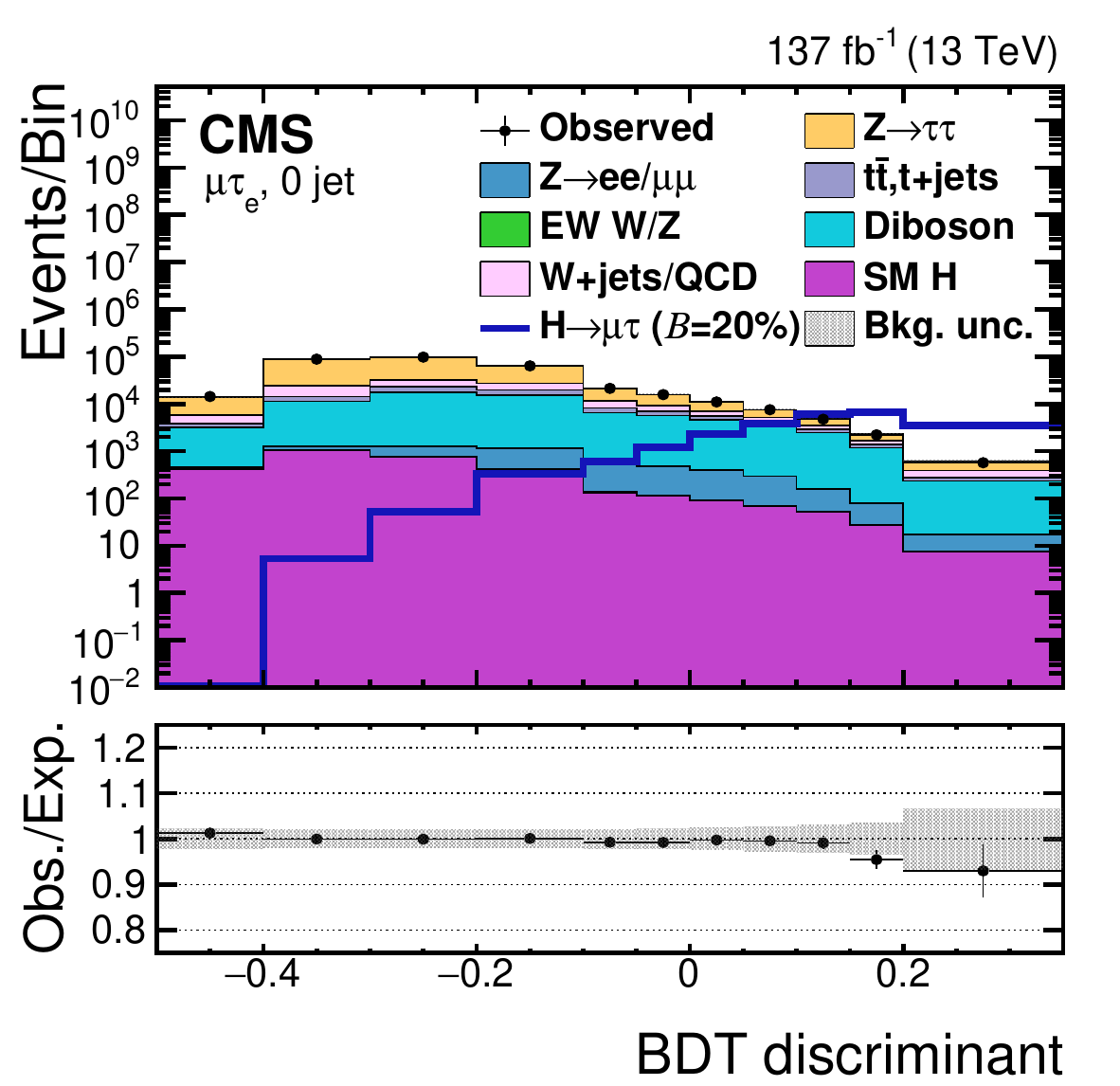}
    \includegraphics[width=2.2in]{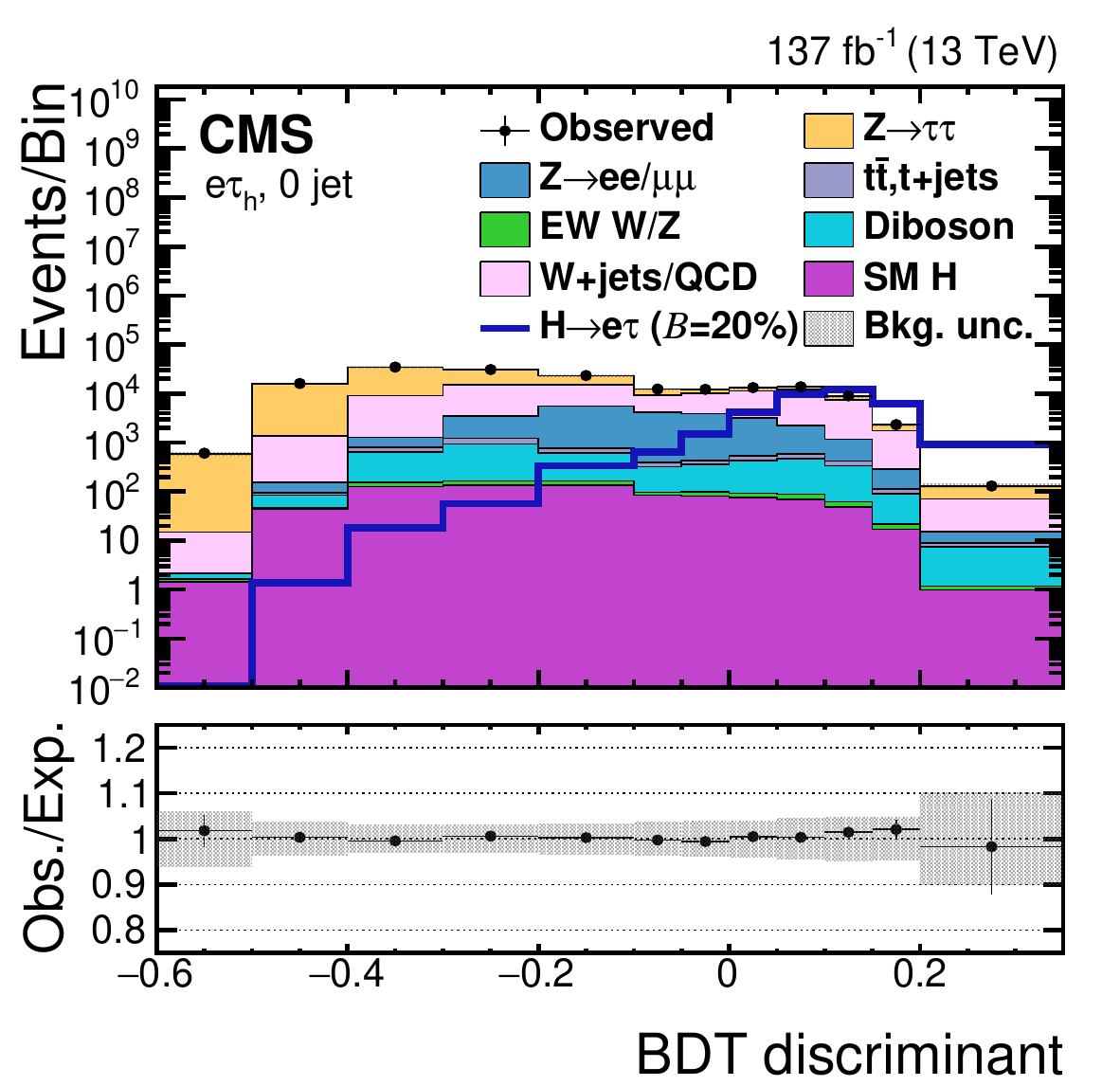}
    \includegraphics[width=2.2in]{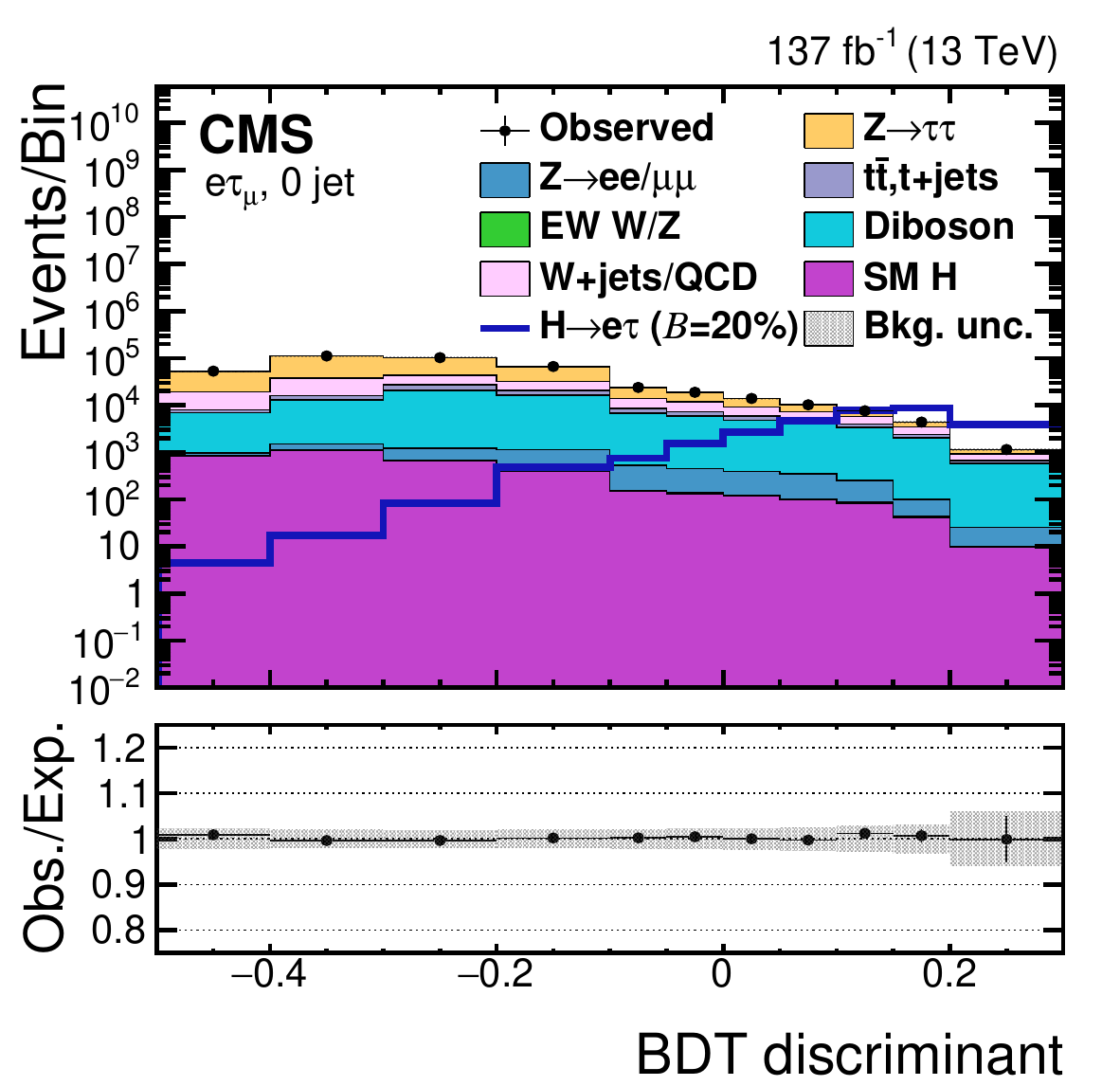}
\vspace*{8pt}
\caption{BDT discrimant distributions for the data and background processes in the 
$H \to \mu \tau_h$ (upper left),  $H \to \mu \tau_e$ (upper right),
$H \to e \tau_h$ (lower left), and $H \to e \tau_\mu$ (lower right) channels.
Only the 0-jet category of each final state is shown here.
A $\mathcal{B}(H \to \mu \tau$) or $\mathcal{B}(H \to e \tau$) = 20\% is assumed for the signal. 
The lower panel in each plot shows the ratio of data and estimated background.
The uncertainty band corresponds to the background uncertainty in which the post-fit statistical and
systematic uncertainties are added in quadrature.
\protect\label{higgs_bdt}}
\end{figure}

BDT is trained to further distinguish signal from background in each final state ($\mu\tau_h$, $\mu \tau_e$, $e\tau_h$, $e\tau_\mu$). Input observables to the BDT include: \pt($l$), \pt($\tau_h$), $m_{col}$, $p_T^{miss}$, $\Delta\phi(l,\tau_h)$, $\Delta\phi(e,\mu)$, etc.
BDT output distributions for the observed data, the hypothetical signal, and the background are shown in Fig.~\ref{higgs_bdt} (taken from Ref.~\refcite{CMS:2021rsq}). Background is normalised to the best fit values from a signal plus background fit.

The branching fractions of $H \to \mu \tau$ and $H \to e \tau$ are extracted by binned maximum likelihood fit of the BDT output distributions.
No significant signal is observed. The upper limit on $\mathcal{B}(H \to \mu \tau)$ at 95\% CL is set to be 0.15\% with an expected limit 0.15\%; the upper limit on $\mathcal{B}(H \to e \tau)$ at 95\% CL is set to be 0.22\% with an expected limit 0.16\%.

\begin{figure}[hbtp]
    \centering
    \includegraphics[width=2.2in]{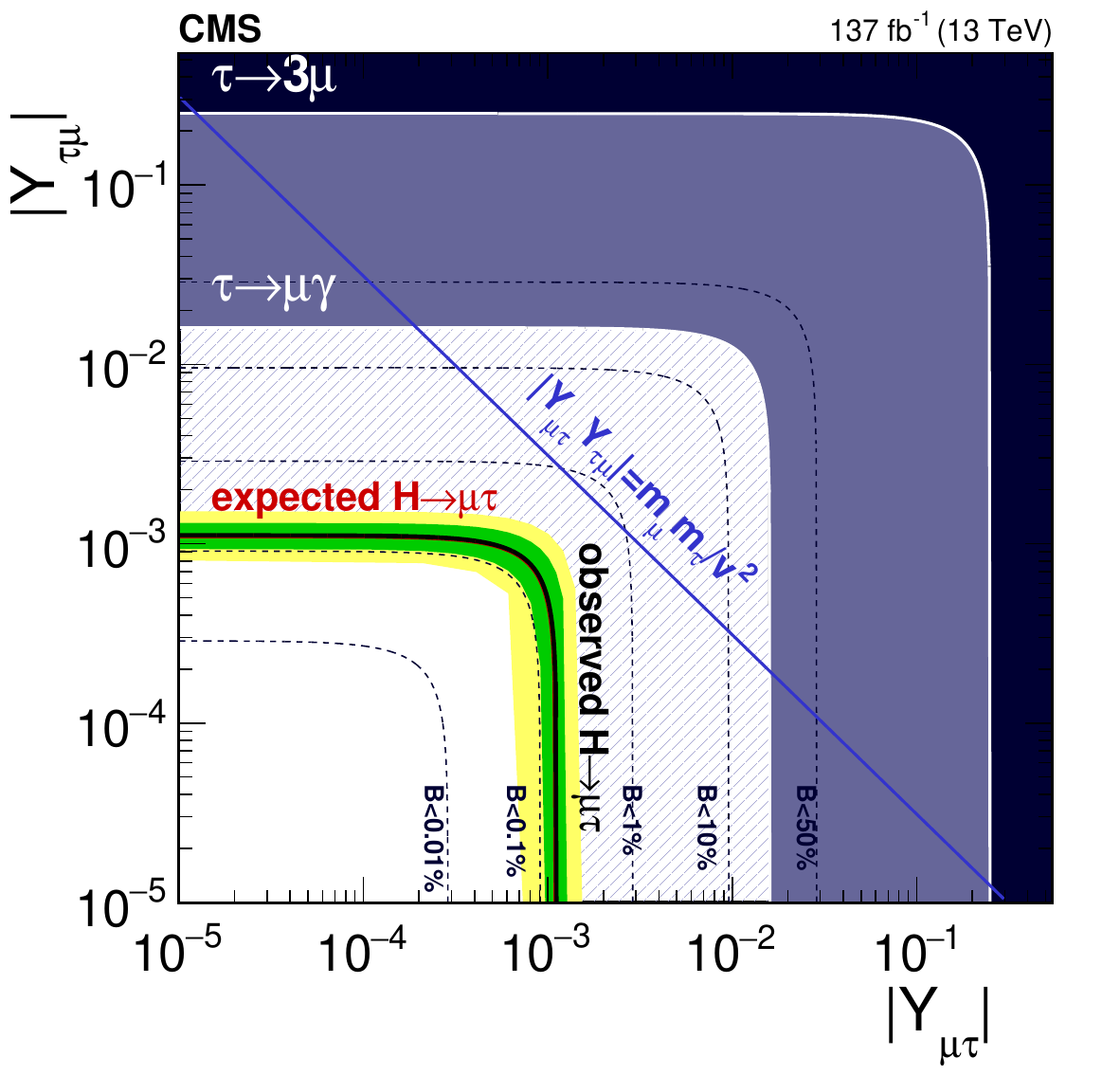}
    \includegraphics[width=2.2in]{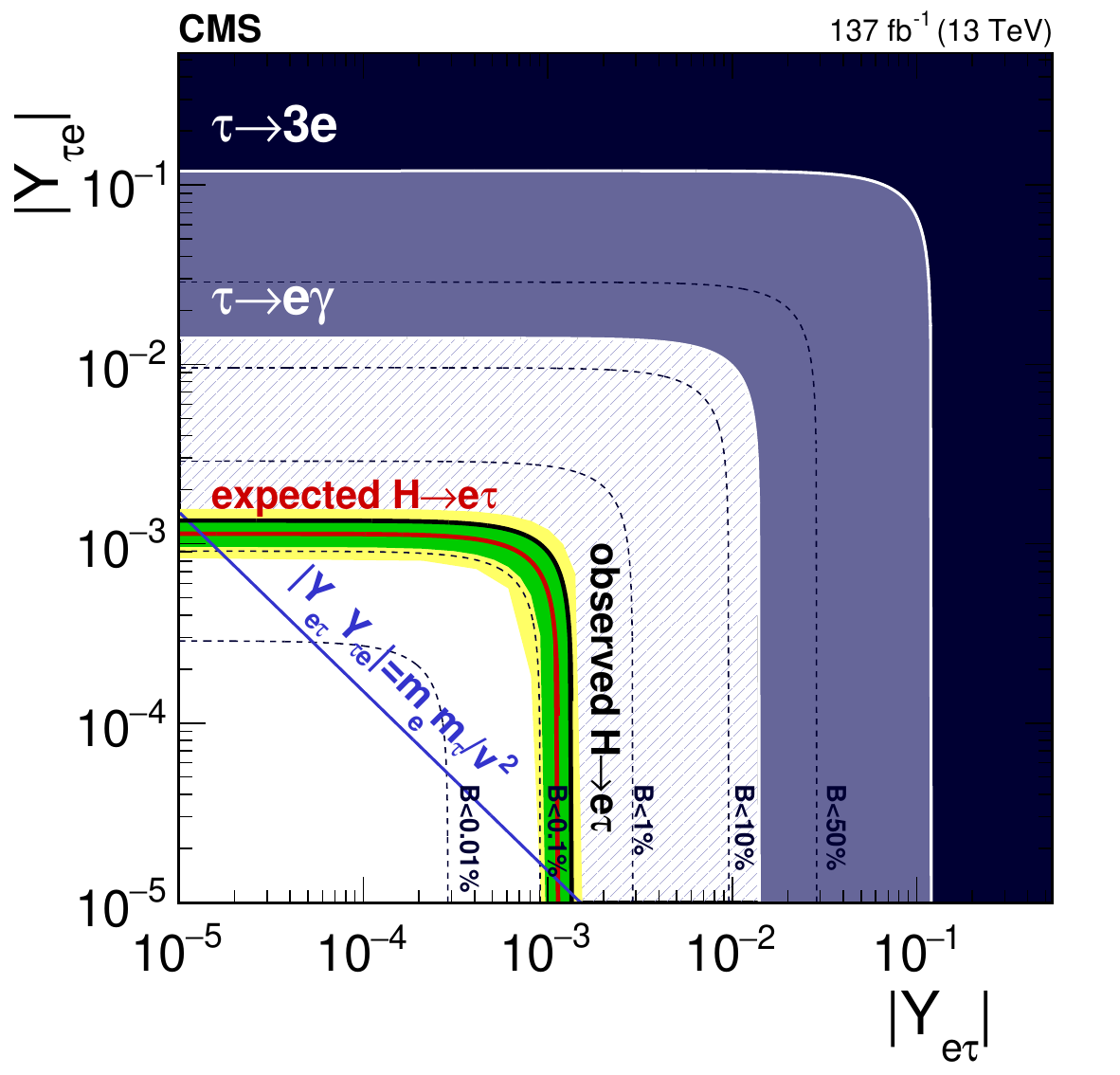}
\vspace*{8pt}
\caption{Expected (red line) and observed (black solid line) 95\% CL upper limits on the LFV Yukawa couplings. In the left plot, the expected limit is covered by the observed limit as they have similar values. The flavor diagonal Yukawa couplings are approximated by their SM values. The green and yellow bands indicate the range that is expected to contain 68\% and 95\% of all observed limit variations from the expected limit. The shaded regions are constraints obtained from null searches for $\tau \to3\mu$ or $\tau\to3e$ (dark blue)~\cite{Hayasaka:2010np} and $\tau\to\mu\gamma$ or $\tau\to e\gamma$ (purple)~\cite{Belle:2007qih}. The blue diagonal line is the theoretical naturalness limit $|{Y_{ij}Y_{ji}}| = {m_i}m_j/v^2$~\cite{Harnik:2012pb}.
\protect\label{higgs_Yukawa}}
\end{figure}

The decay width of LFV $H \to \mu \tau$ or $H \to e \tau$ is
  \begin{equation*}
    \Gamma(H \to \ell^{\alpha} \ell^{\beta}) = \frac{m_H}{8\pi}({|Y_{\ell^{\alpha}\ell^{\beta}}|}^2 + {|Y_{\ell^{\beta}\ell^{\alpha}}|}^2),
  \end{equation*}
while the branching fraction of the LFV Higgs boson decay is related to the decay width through
  \begin{equation*}
    \mathcal{B}(H \to \ell^{\alpha} \ell^{\beta}) = \frac{\Gamma(H \to \ell^{\alpha} \ell^{\beta})}{\Gamma(H \to \ell^{\alpha} \ell^{\beta}) + \Gamma_{\mathrm{SM}}}.
  \end{equation*}
Therefore the upper limits on decay branching fractions could be converted to upper limits on Higgs LFV Yukawa interactions, as shown in Figure~\ref{higgs_Yukawa} (taken from Ref.~\refcite{CMS:2021rsq}). 
The results by direct search of LFV $\tau$ decays are superimposed.
Results of branching fractions and Yukawa interactions are also summarized in Table~\ref{result_higgs}.

\begin{table}[htbp!]
\centering
\tbl{Summary of observed and expected upper limits at 95\% CL, best fit branching fractions and corresponding constraints on Yukawa couplings. }
{\begin{tabular}{lccc}
     & Observed (expected) & Best fit branching & Yukawa coupling                \\
     &  upper limits (\%)  &   fractions (\%)   &   constraints                  \\
$H \to \mu \tau$ &   $<$0.15 (0.15)    &   $0.00\pm0.07$    & $<1.11\,(1.10){\times}10^{-3}$ \\
$H \to e \tau$ &   $<$0.22 (0.16)    &   $0.08\pm0.08$    & $<1.35\,(1.14){\times}10^{-3}$ \\
\end{tabular}
\label{result_higgs}}
\end{table}

The ATLAS experiment has published the same analysis and the results are very similar~\cite{ATLAS:2023mvd}.
CMS has also searched for LFV $H \to e\mu$ decay~\cite{CMS:2023pte}, and the observed upper limit on $\mathcal{B}(H \to e\mu)$ is determined to be $4.4 \times 10^{-5}$ at 95\% CL - this branching fraction is strongly constrained by the $\mu \to e\gamma$ limit to be $\mathcal{B}(H \to e\mu)< 10^{-9}$, though.

\section{Search for LFV in the top quark sector}
\label{sec:top}

\begin{figure}[hbtp]
\centerline{
\includegraphics[width=5in]{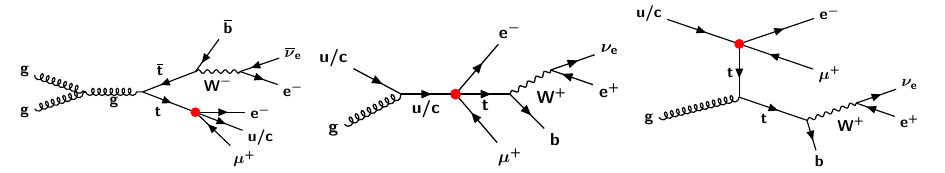}
}
\vspace*{8pt}
\caption{Feynman diagrams for top quark decays in $t\bar{t}$ events (left) and single top quark production (middle and right) via LFV interactions.
The LFV interaction vertex is shown as a solid red circle to indicate that it is not allowed in the SM.
\protect\label{top_diagram}}
\end{figure}

The top quark, with a mass of about 173 GeV, is the heaviest elementary particle in the SM. The LHC is practically a top quark factory, with more than $10^8$ top quarks produced over the whole Run 2, offering an opportunity to study LFV processes involving top quark production or decays.
The signal processes considered in the CMS searches for LFV in the top quark sector include single top production as well as decays of top quarks in $t\bar{t}$ production, as shown in Fig.~\ref{top_diagram}.
Depending on how the W boson decays ($qq$ or $l\nu$), the final state could be either dilepton ($e\mu$) or trilepton ($e\mu e$ or $e\mu\mu$). Both final states have been explored by the CMS experiment~\cite{CMS:2022ztx,CMS:2023iul}, while the trilepton analysis is detailed here for its significantly better sensitivity.

Th LFV signals are parameterized using dimension-6 effective field theory (EFT) operators:
\begin{equation}
\mathcal{L}=\mathcal{L}_{\text{SM}}^{(4)} + \frac{1}{\Lambda^2}\sum_{a}C_{a}^{(6)}O_{a}^{(6)}+O\left( \frac{1}{\Lambda^4}\right),
\end{equation}  
where $\mathcal{L}_{\text{SM}}^{(4)}$ is the renormalizable Lagrangian of the SM. The $O_{a}^{(6)}$ term denotes dimension-6 nonrenormalizable operators, and the $C_{a}^{(6)}$ terms are the corresponding Wilson coefficients. The dimension-6 terms are suppressed by the square of a mass scale $\Lambda$ where new physics is presumed to emerge.
The results are interpreted in terms of limits on vector, scalar, and tensor interactions originating from dimension-6 operators within the EFT framework. Details of the framework can be found in Ref.~\refcite{Durieux:2014xla,Aguilar-Saavedra:2018ksv}

Data events are recorded using a combination of single-, double-, and triple-lepton triggers.
The offline selection requires one opposite-charge electron-muon pair, a third charged lepton (either electron or muon), at least one jet, of which no more than one is tagged as b-jet,
and $p^{miss}_T$ larger than 20~GeV to target the neutrino from the W boson decay.
Events with an opposite-charge, same-flavor lepton pair of invariant mass between 50 and 106~GeV are discarded ("Z veto").

Background processes with three genuine leptons (dominated by WZ+jets) are taken from simulation. The data and simulation prediction are compared in a control region, defined by inverting the "Z veto", i.e. requiring one pair of same-flavor but opposite-charge leptons having invariant mass between 50 and 106~GeV, and with at least 1 jet, of which at most one is b-tagged. This control region exhibits a good agreement between data and simulation. 
The background with two genuine leptons and one misidentified, including  $t\bar{t}$ and Drell-Yan processes, is estimated using a Matrix Method~\cite{Gillam:2014xua} from data.

The top quark decay signal (Fig.~\ref{top_diagram}, left) and the top quark production signal (Fig.~\ref{top_diagram}, middle and right) differ in kinematics. Notably the invariant mass of the electron and muon, $m(e\mu)$, tends to be large due to the presence of high \pt lepton(s) in the top quark production events, while it is bounded by the top quark mass in the top quark decay events. The selected events are therefore divided into two signal regions (SRs): 
\begin{itemlist}
 \item SR1: $m(e\mu) < 150$~GeV, targeting top quark decay signal;
 \item SR2: $m(e\mu) > 150$~GeV, targeting top quark production signal.
\end{itemlist}

\begin{figure}[hbtp]
    \centering
    \includegraphics[width=2.2in]{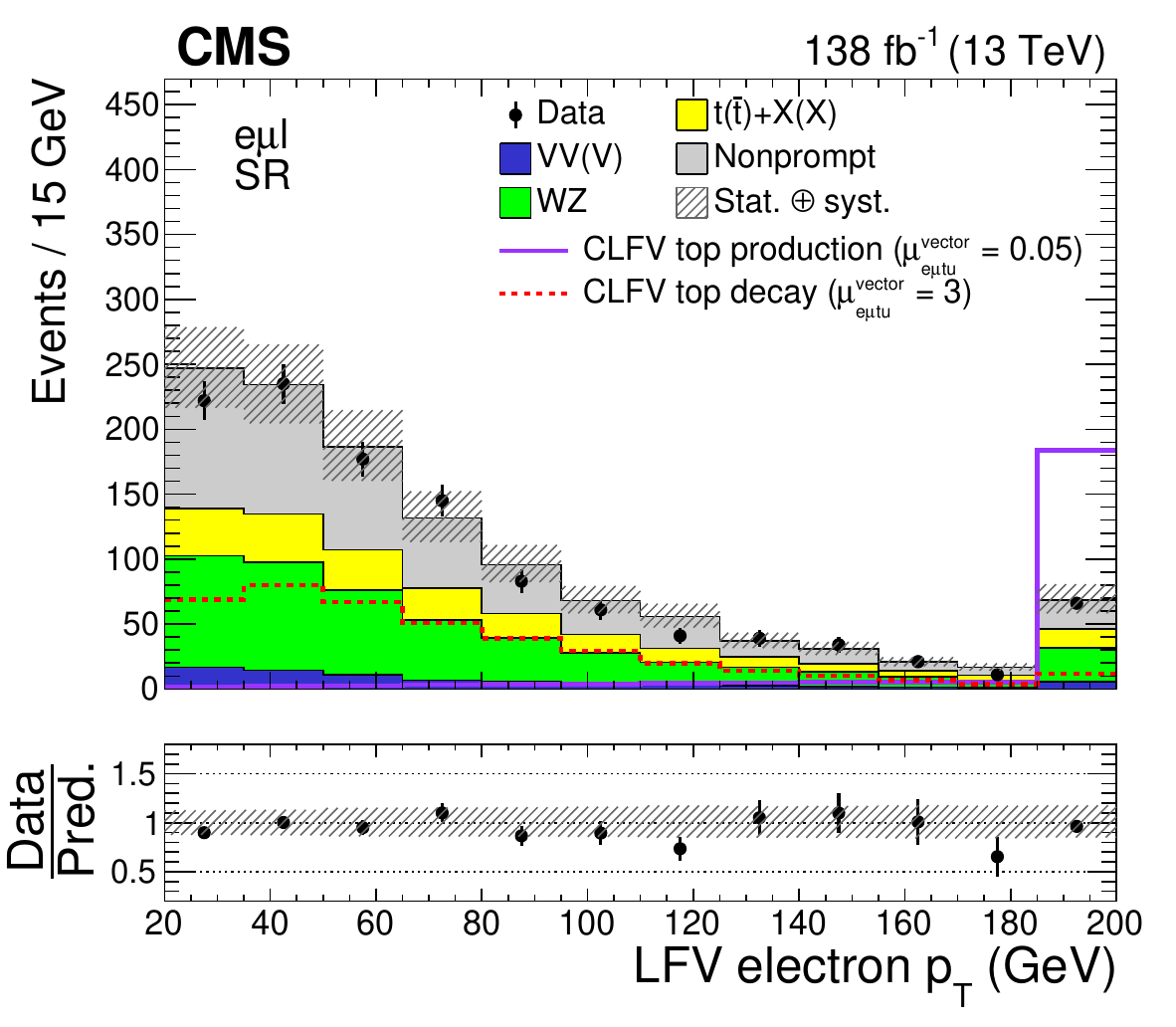}
    \includegraphics[width=2.2in]{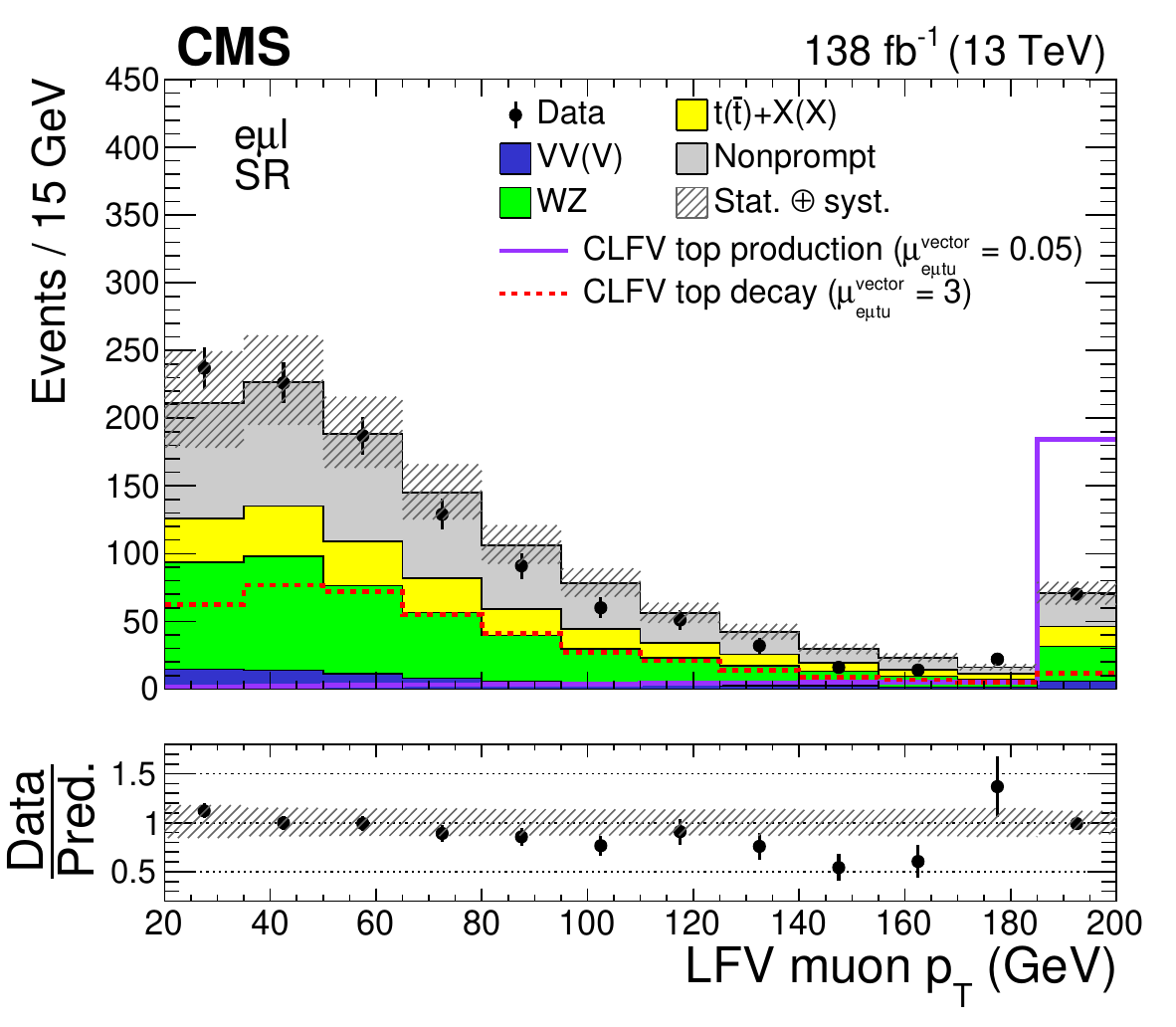}
    \includegraphics[width=2.2in]{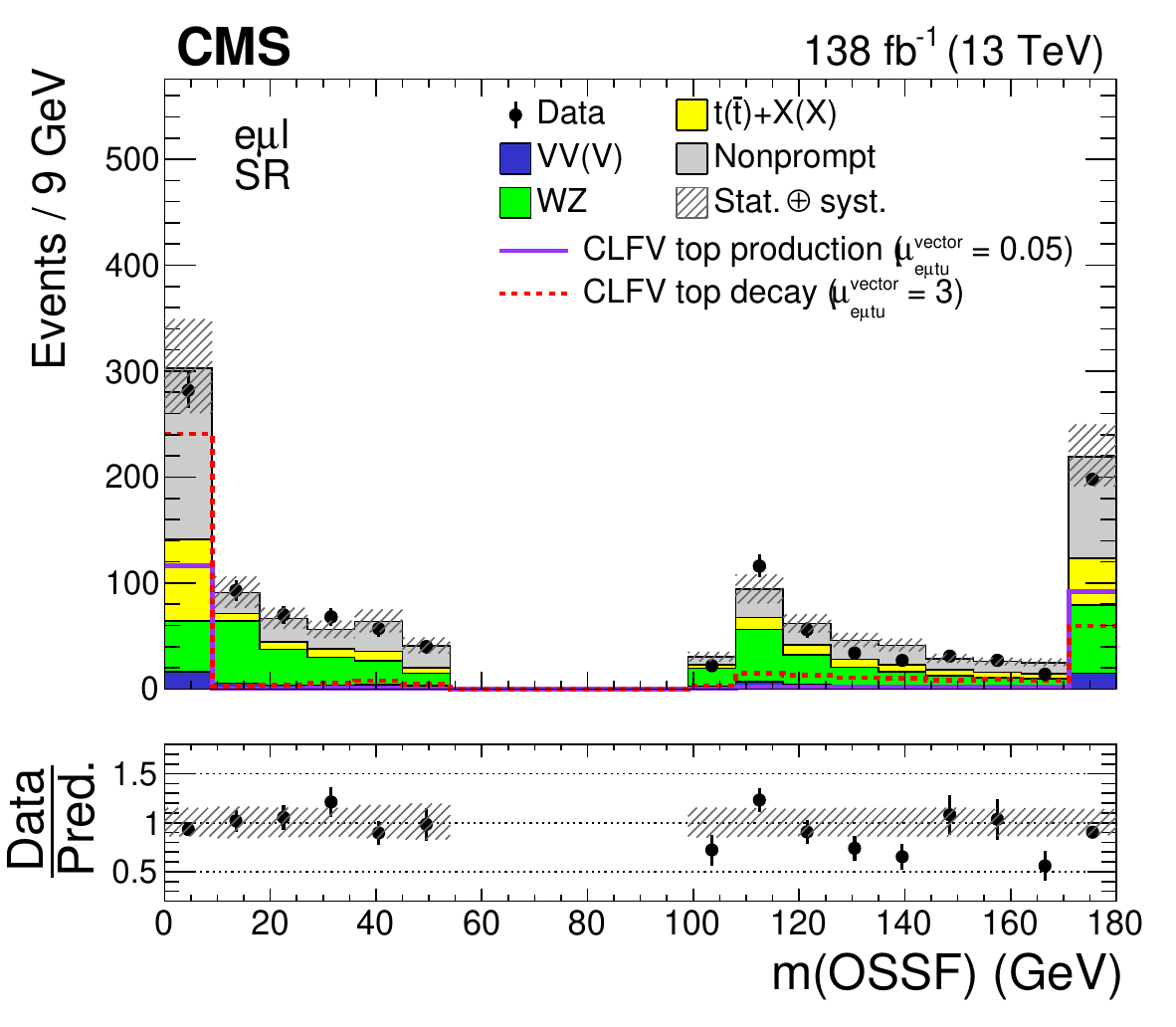}
    \includegraphics[width=2.2in]{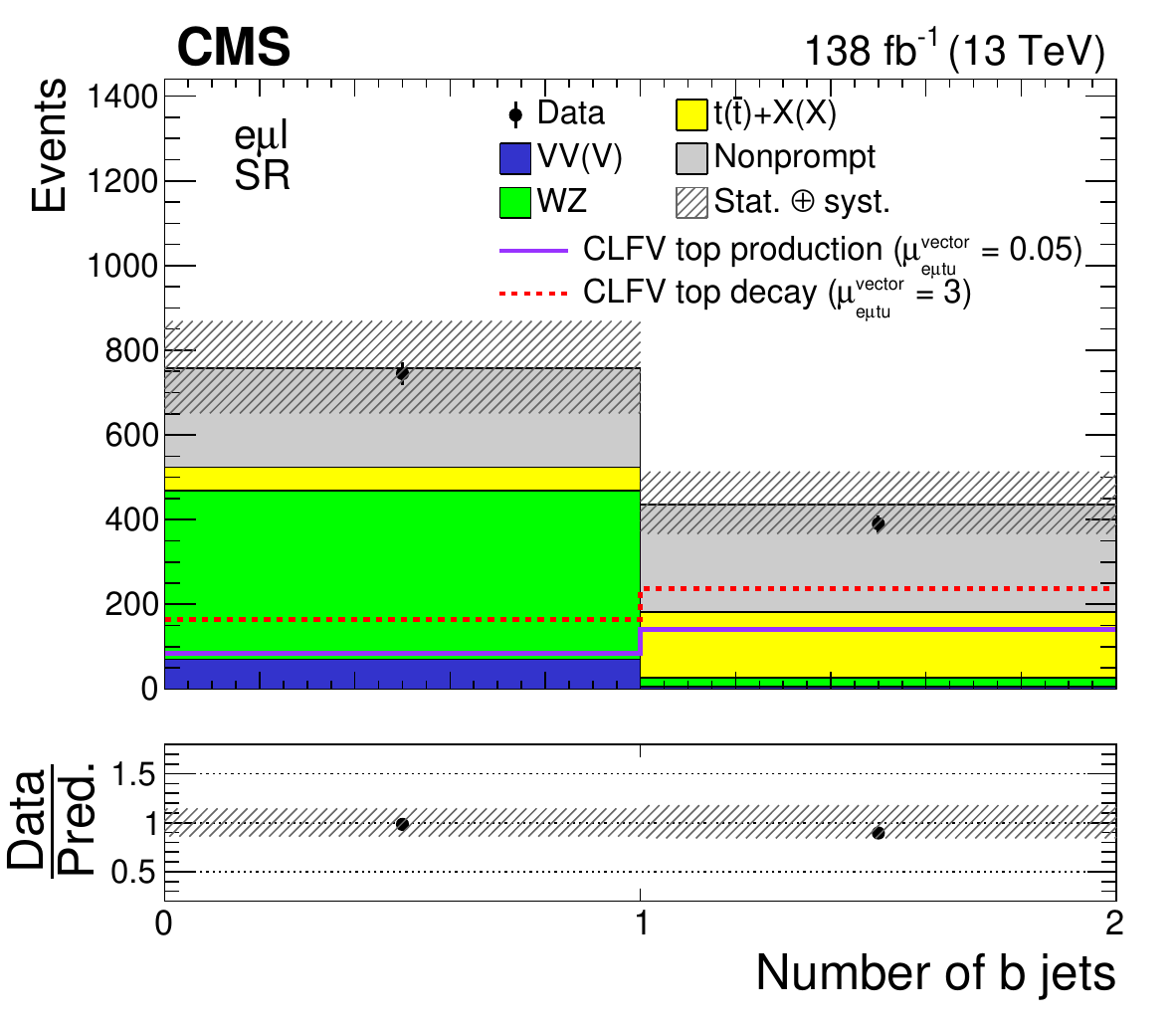}
\vspace*{8pt}
\caption{Distributions of kinematic variables in the SR: LFV electron \pt (upper left), LFV muon \pt (upper right), opposite-charge same-flavor lepton pair mass (lower left), and $b$ jet multiplicity (lower right). The LFV top quark decay and production signals are shown as dotted red and solid purple lines, respectively. The original signal normalization, corresponding to $C_{e\mu tu}^{\text{vector}}/\Lambda^2=1~\text{TeV}^{-2}$, is scaled up (down) by a factor of 3 (20) for the LFV top quark decay (production) signal for better visualization. The hatched bands indicate statistical and systematic uncertainties in the background predictions. The last bin of all but the lower-right histogram includes the overflow events.
\protect\label{top_inputs}}
\end{figure}

\begin{figure}[hbtp]
    \centering
    \includegraphics[width=2.2in]{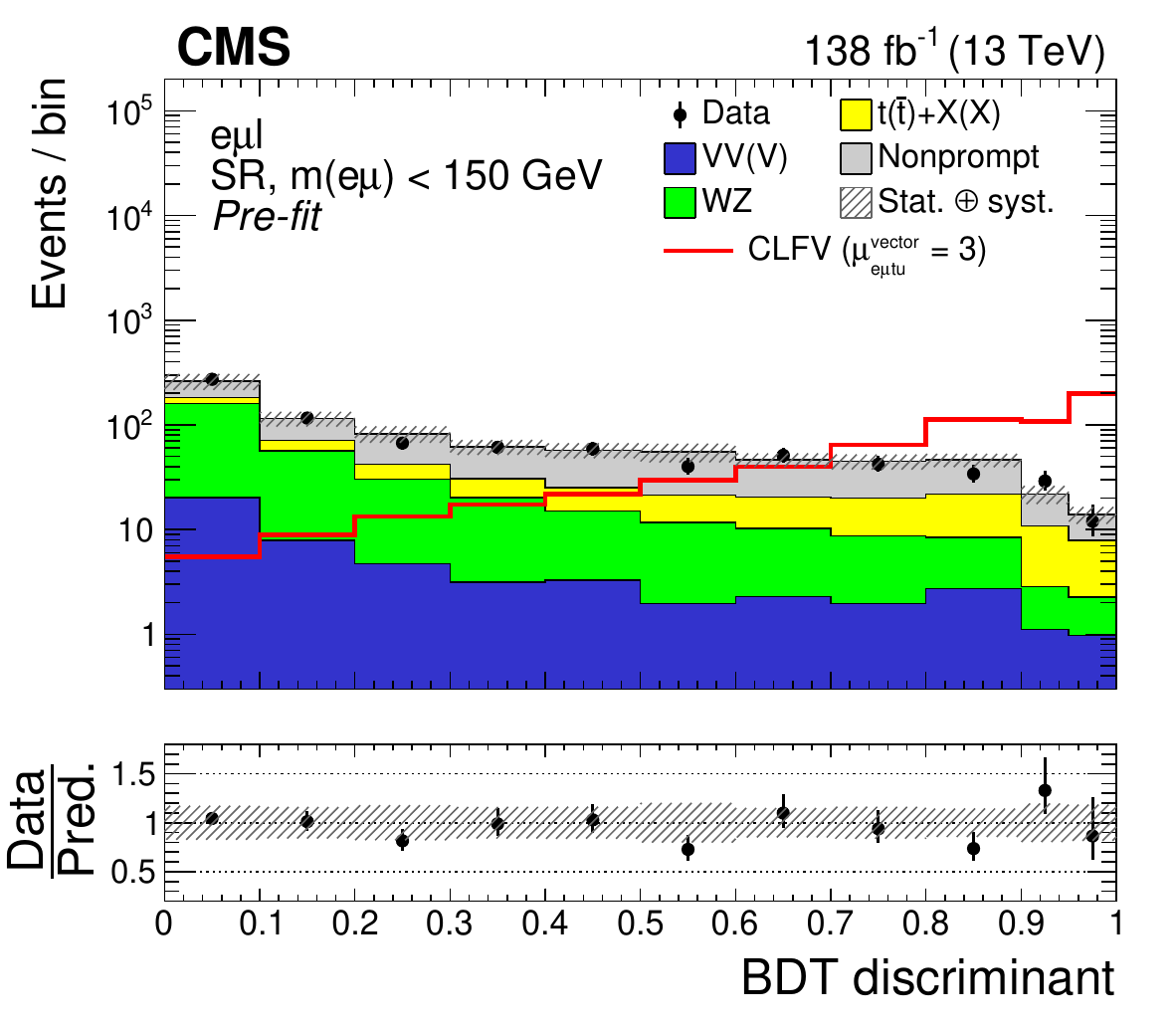}
    \includegraphics[width=2.2in]{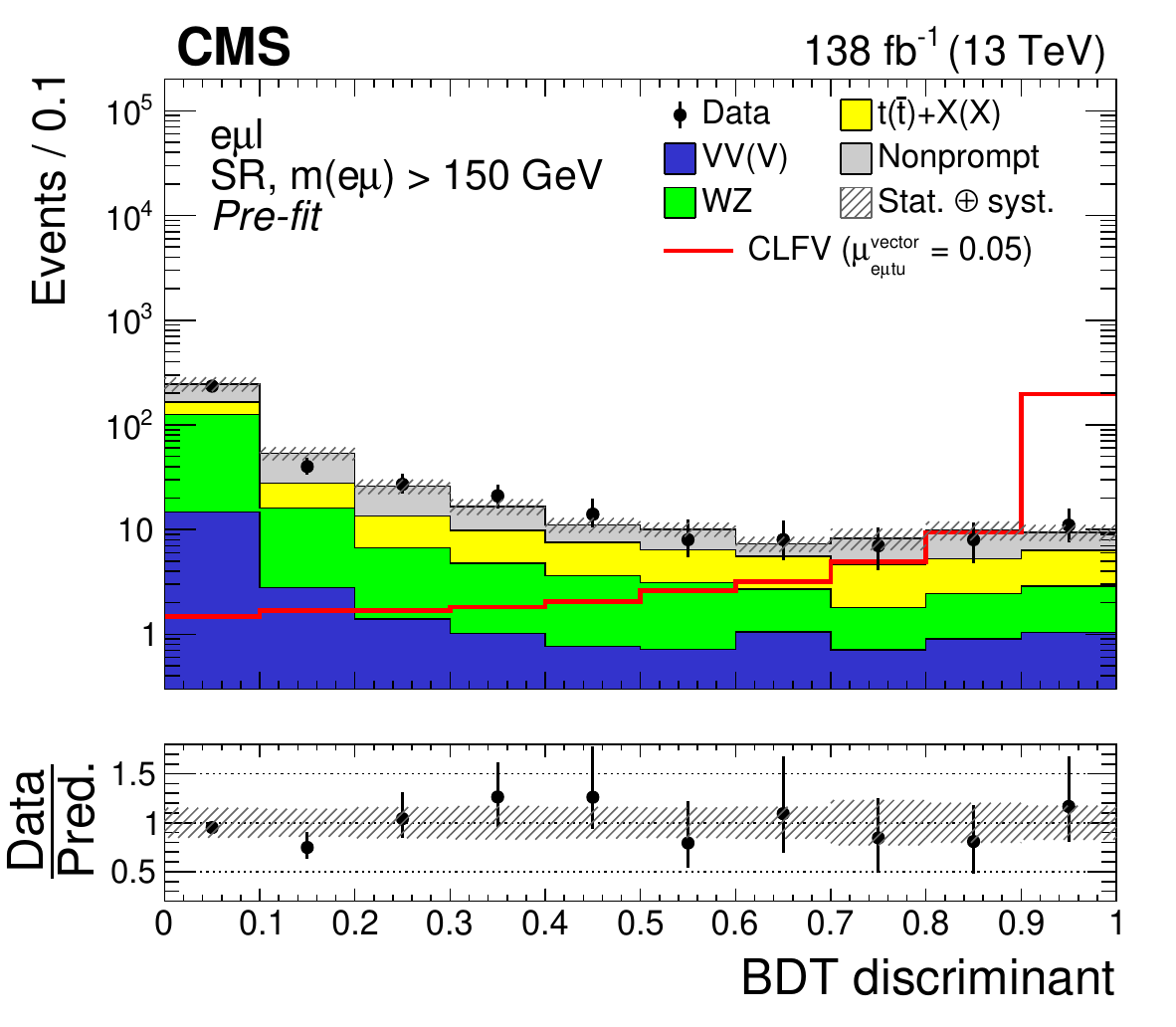}
\vspace*{8pt}
\caption{Distributions of the BDT discriminant targeting the LFV top quark decay (left) and production (right) signal. Contributions from the two signal modes (production and decay) are combined within each SR and are shown as the solid red line. The pre-fit signal strength ($\mu_{e\mu tu}^{\text{vector}}=1$), corresponding to $C_{e\mu tu}^{\text{vector}}/\Lambda^2=1~\text{TeV}^{-2}$, is scaled up (down) by a factor of 3 (20) for the LFV top quark decay (production) signal for better visualization. The hatched bands indicate statistical and systematic uncertainties in the background predictions.
\protect\label{top_outputs}}
\end{figure}

One BDT in each SR is trained, taking kinematic observables of the final state particles as inputs, as shown in Fig.~\ref{top_inputs} (taken from Ref.~\refcite{CMS:2023iul}). The BDT outputs of data, background expectation, and hypothetical signal, are shown in Fig.~\ref{top_outputs} (taken from Ref.~\refcite{CMS:2023iul}). The total expected backgrounds are in good agreement with the data observation. SR2 has a significantly better signal-to-background ratio.

A simultaneous binned maximum likelihood fit is performed on the six BDT ouput distributions (three data-taking years and two SRs). Upper limits are set on the Wilson coefficients and branching fraction, as summarised in Table~\ref{top_results}. The upper limit on a certain Wilson coefficient is obtained by setting other Wilson coefficients to zero. Upper limits on branching fractions are obtained assuming the energy scale is 1~TeV. The different sensitivities among scalar, vector and tensor are due to signal acceptance. 

\begin{table*}[tbh]
\centering
\tbl{Upper limits at 95\% CL on Wilson coefficients and the branching fractions for tensor-, vector-, and scalar-like LFV interactions. The expected and observed upper limits are shown in regular and bold fonts, respectively. The intervals that contain 68\% of the distribution of the expected upper limits are shown in parentheses.}
{\begin{tabular}{cccccc}
LFV       & Lorentz  & \multicolumn{2}{c}{$C_{e\mu tq}/\Lambda^2~(\text{TeV}^{-2})$} & \multicolumn{2}{c}{$\mathcal{B} (t \to e\mu q) \times 10^{-6}$} \\
coupling  & structure & Exp. (68\% CL range) & \textbf{Obs.} & Exp. (68\% CL range) & \textbf{Obs.} \\
\noalign{\vskip 1mm}
\hline
\noalign{\vskip 1mm}
\multirow{3}{*}{$e\mu tu$}& Tensor & 0.022 (0.018--0.026) & \textbf{0.024} & 0.027 (0.018--0.040) & \textbf{0.032}\\
& Vector & 0.044 (0.036--0.054) & \textbf{0.048} & 0.019 (0.013--0.028) & \textbf{0.022}\\
& Scalar & 0.093 (0.077--0.114) & \textbf{0.101} & 0.010 (0.007--0.016) & \textbf{0.012}\\
\noalign{\vskip 1mm}
\hline
\noalign{\vskip 1mm}
\multirow{3}{*}{$e\mu tc$} & Tensor & 0.084 (0.069--0.102) & \textbf{0.094} & 0.396 (0.272--0.585) & \textbf{0.498}\\
 & Vector & 0.175 (0.145--0.214) & \textbf{0.196} & 0.296 (0.203--0.440) & \textbf{0.369}\\
 & Scalar & 0.385 (0.318--0.471) & \textbf{0.424} & 0.178 (0.122--0.266) & \textbf{0.216}\\
 \noalign{\vskip 1mm}
\hline
\end{tabular}\label{top_results} }
\end{table*}

\section{Summary}
\label{sec:summary}

Unlike quark mixing and neutrino mixing, charged lepton flavor violation (LFV) has never been observed, despite a large number of experiments looking for it over the decades. A wide range of processes could be explored in searching for LFV, and the relations of various LFV decay modes are model dependent. Therefore, the LFV searches performed at CMS, a high-energy proton-proton collision experiment, are complementary to those at low-energy high-intensity muon beams or electron-position colliders.
In this review we have reported LFV searches in the \ttm~decay, in the Higgs boson decays, and in the top quark production and decays. The data correspond to an integrated luminosity of up to $138~fb^{-1}$ of proton-proton collisions at a center-of-mass energy of 13~TeV taken during 2016-2018.
The observed upper limit on the branching fraction of $\tau\to 3\mu$ is $2.9 \times 10^{-8}$ at 90\% confidence level (CL).
The observed upper limits on the branching fraction of $H \to \mu \tau$ are 0.15\%, and of $H \to e \tau$ are 0.22\%, respectively, at 95\% CL.
The observed upper limits on the branching fractions of $t\to e\mu u$ ($t\to e \mu c$) are $0.032 (0.498) \times 10^{-6}$, $0.022 (0.369) \times 10^{-6}$, and $0.012 (0.216) \times 10^{-6}$ for tensor, vector, and scalar interactions, respectively, at 95\% CL.
These null results allow stringent constraints to be placed on theories beyond the Standard Model.
As most of these searches are dominated by statistical uncertainties, the additional Run 3 data being collected will lead to further advances in these crucial tests of the Standard Model.

\bibliographystyle{ws-mpla}
\bibliography{ws-mpla}

\end{document}